\documentclass[fleqn,usenatbib]{mnras}

\usepackage{newtxtext,newtxmath}
\usepackage[T1]{fontenc}

\usepackage{graphicx}
\usepackage{amsmath}
\usepackage{multirow}

\newcommand{\swift}{\textit{Swift}}
\newcommand{\eb}{$\epsilon_{B}$ }

\title[LOFAR prompt follow-up of GRBs]{A LOFAR search for coherent radio emission accompanying prompt engine activity in gamma-ray bursts}

\author[A. Hennessy et al.]{A. Hennessy$^{1}$,\thanks{E-mail: ah724@leicester.ac.uk}
R. L. C. Starling$^{1}$,
A. Rowlinson$^{2,3}$,
I. de Ruiter$^{2}$,
A. J. van der Horst$^{4}$,
G. E. Anderson$^{5}$, \newauthor
N. R. Tanvir$^{1}$,
S. ter Veen$^{3}$,
K. Wiersema$^{6}$,
R. A. M. J. Wijers$^{2}$\\
\\
$^{1}$School of Physics and Astronomy, University of Leicester, University Road, Leicester LE1 7RH, UK\\
$^{2}$Anton Pannekoek Institute for Astronomy, University of Amsterdam, Science Park 904, 1098 XH Amsterdam, The Netherlands\\
$^{3}$ASTRON, the Netherlands Institute for Radio Astronomy, Oude Hoogeveensedijk 4, 7991 PD Dwingeloo, The Netherlands\\
$^{4}$Department of Physics, George Washington University, 725 21st St NW, Washington, DC 20052, USA\\
$^{5}$International Centre for Radio Astronomy Research, Curtin University, GPO Box U1987, Perth, WA 6845, Australia\\
$^{6}$Centre for Astrophysics Research, University of Hertfordshire, Hatfield, AL10 9AB, UK\\
}

\date{Accepted XXX. Received YYY; in original form ZZZ}

\pubyear{2025}

\begin{document}
\label{firstpage}
\pagerange{\pageref{firstpage}--\pageref{lastpage}}
\maketitle

\begin{abstract}
    Relativistic jets generated in gamma-ray bursts (GRBs) produce luminous transient events, yet the fundamentals of jet composition and radiation mechanisms remain unclear. One means of identifying a magnetically-dominated outflow would be detection of prompt, coherent radio emission at low frequencies, and we are able to search for this using the LOw Frequency ARray (LOFAR) coupled with modelling of high energy pulses detected by the Neil Gehrels Swift Observatory (\swift). We present the rapid response mode follow-up LOFAR observations of four long GRBs, each beginning within a few hundred seconds of the initial \swift-BAT trigger. We interpreted our findings under the framework of a magnetic wind model, predicting coherent radio emission analogous to prompt emission pulses. Using 60 second and 180 second time sliced imaging at 120--168 MHz, we obtain upper limits on radio pulse emission, finding no significant signals. In the case of GRB\,200925B, we observed a small increase of radio flux seen at $\sim$60--360\,s post burst. In this model, this could represent the radio emission related to the \swift-BAT pulses, for a redshift of $z=1.8$, however, with a low signal-to-noise ratio of $\sim 2$, it is not deemed significant enough to confirm coincident prompt radio and gamma-ray emission. Instead, we can constrain the \eb parameter, deriving upper limits of $\epsilon_{B} < 4.2 \times 10^{-4}$ for GRB\,200925B. In GRB\,240414A, with a reported redshift of $z=1.833$, we constrain $\epsilon_{B} < 2.8 \times 10^{-4}$. We discuss these results in the context of our whole LOFAR rapid response sample of six long gamma-ray bursts, finding our \eb values are generally consistent with previous GRB studies.
\end{abstract}


%%%%%%%%%%%%%%%%% BODY OF PAPER %%%%%%%%%%%%%%%%%%

\section{Introduction}
    \label{sec:introduction}

    Gamma-ray bursts (GRBs) are the most energetic class of high energy transients seen in the Universe, powered by a highly collimated relativistic jet. Since the first detection in 1967 \citep{klebesadel1973}, thousands of GRBs have been observed yet many aspects of GRB physics remain poorly understood.
    
    In this paper, we focus on the most commonly detected subset of GRBs known as long gamma-ray bursts (LGRBs), making up approximately 85 per cent \citep{lien2016} of bursts observed with the \textit{Neil Gehrels Swift Observatory} \citep[hereafter \swift,][]{gehrels2004}. Long GRBs occur during the deaths of a massive star in a core-collapse supernova. They typically occur with durations longer than $\sim2$\,s \citep{kouveliotou1993}, but this is detector dependent, and there is some overlap with the population of short gamma-ray bursts (SGRBs).

    A GRB is characterised observationally by two key phases - the initial prompt high energy emission generated within the relativistic jet, and longer lasting broadband afterglow emission as the relativistic jet interacts with its environment. The prompt emission is observed as a number of highly energetic gamma-ray pulses on timescales of a few ms to a few hundred seconds, likely caused by the interaction of internal energy shocks within the relativistic jet resulting in the behaviour observed \citep{sari1997a}. However, the nature of emission and the distance from the central engine that these shocks occur are still debated, in part due to the unknown composition of the jet. Additionally, a thermal component from the photosphere may be a significant contributor to emission in at least some bursts \citep{peer2007,ryde2010}. A Poynting flux dominated jet can produce the prompt emission through a magnetic reconnection \citep[e.g.][]{usov1992}. This works at higher efficiencies, but such models are highly dependent on magnetic field strength and configuration within the jet which is not well known. \citet{rees1994} and \citet{lazzati2013}, for example, favour a matter-dominated jet, where collisions between shells at different Lorentz factors produce synchrotron radiation - one issue with this model is the relatively low efficiency of emission being in tension with energetics of the X-ray afterglow observations. 

    The subsequent afterglow displays temporal and spectral behaviour typical of synchrotron emission. X-ray afterglows are detected in the majority of GRBs \citep{evans2009}, and afterglows are also seen at UV/optical through to radio wavelengths \citep[e.g.][]{kann2011,anderson2018}. The detection of radio emission has proven a key diagnostic for the synchrotron-dominated afterglow physics \citep{waxman1998,chandra2012}; and jet geometry, structure and energetics \citep[e.g.][]{horst2008,granot2014,rhodes2024}. 
    
    Radio signatures are also expected during the prompt emission phase for some magnetically dominated models. Magnetic reconnection is capable of releasing quick, powerful bursts that can reproduce the variability of GRB light curves \citep{beniamini2016}. At least one model predicts low frequency radio emission to be emitted alongside the higher energy component through theoretical and simulation studies \citep{usov2000,smolsky2000}. In such models, the shock front of a relativistic, magnetized wind interacts with the ambient media, creating a surface current at the boundary. The strong electric field here accelerates electrons to relativistic speeds, causing X-ray and gamma-ray emission through Synchrotron radiation. The interaction is highly variable, and the varying electric field induces low-frequency electromagnetic radiation.
    
    This model potentially offers us a testable prediction for magnetically dominated GRB outflow: a radio flare analogous to gamma-ray pulses should be emitted, but observed later due to a dispersion delay \citep{taylor1993}. As radiation travels through the host galaxy environment, intergalactic medium and the Milky Way, longer wavelengths undergo greater dispersion, causing radio emission to arrive with a delay compared to the high energy regime, typically on the order of a few hundred seconds at low radio frequencies. The brightness of the resultant radio flare depends on the energetics of the gamma-ray pulse, these are commonly modelled with fast-rise exponential-decay peaks \citep[FREDs,][]{norris1996} which can then be integrated across to calculate the total fluence.
    
    The peak frequency of the emission in the magnetic wind models of \citet{usov2000} lies well below the capabilities of current radio facilities at $\sim0.01$\,MHz, but telescopes like the LOw Frequency ARray \citep[LOFAR,][]{vanhaarlem2013} with recievers at 10-80\,MHz and 120-240\,MHz; and the Murchison Widefield Array \citep[MWA,][]{tingay2013}, operating at 80-300\,MHz, should detect the higher frequency tail of this emission. \citet{starling2020} provides a basis to show how LOFAR is capable of observing radio flares within this model, and establishing the mechanism for prompt GRB emission if found. This was done by demonstrating the detectability of predicted radio flares with LOFAR, given timing and flux density predictions calculated from a \swift-XRT flare sample with the magnetic wind model applied. While the study leads us to expect a significant fraction of GRB X-ray flares would be detectable with LOFAR High Band Antenna (HBA), a number of assumptions are folded into their calculations, and we re-examine these following the results of our study.
    
    The MWA rapid-response mode \citep{hancock2019} is capable of repointing within 10\,s of receiving a \swift\ GRB alert. Combined with dispersion delay, this makes MWA rapid enough to constrain prompt radio signals associated with the shocks within the jet. \citep{anderson2021,tian2022} as well as X-ray flares \citep{tian2022a}. Responsive modes have also been developed for the Long Wavelength Array at Owens Valley Radio Observatory to capture fast radio transients associated with GRBs and gravitational waves \citep{kosogorov2025}. While LOFAR's current response time of $\sim$4--5 minutes means it will not catch the very first radio pulses from a GRB, its superior sensitivity allows us to deeply probe radio pulses associated with X-ray flares \citep{hennessy2023} and other prompt radio emission mechanisms \citep{rowlinson2019,rowlinson2021,rowlinson2024}.

    One of the more uncertain parameters is the fraction of total energy in the magnetic field in the GRB jet, $\epsilon_{B}$. This has direct consequences on interactions within the jet, and importantly it controls the brightness of radio emission in the magnetic wind model. The dynamic nature of the event and surroundings means this parameter should evolve over the lifetime of the burst \citep{ror2024} and is likely somewhat variable across different bursts. Several studies obtain \eb value estimations through GRB afterglows \citep[e.g.][]{barniolduran2014,beniamini2015,leung2021}, finding a wide range of values and upper limits spanning many orders of magnitude. However, the dynamic evolution of the jet, and hence also the magnetic field of the jet, means that \eb may not be directly comparable between the prompt, as studied in this paper, and afterglow phases.

    Especially during the prompt emission phase, the transient nature of GRBs can make detection and follow-up inconsistent. Each burst is unique temporally and spectrally, and while many show common features, no model has been successful in explaining all observations - see \citet{kumar2015} for a comprehensive review. Prompt emission coverage has been challenging at low frequencies, however, two LGRBs have already been investigated through a rapid response follow-up with LOFAR \citep{rowlinson2019,hennessy2023} and a further one through follow-up with MWA \citep{tian2022a}. All of these early-time studies draw similar conclusions: no coherent radio emission is detected allowing for constraints to be placed on the parameters within the magnetic reconnection model of \citet{usov2000} - in particular \eb, estimating upper limits of $\sim 10^{-4}$.

    This only represents a total of three attempts at observing prompt radio emission from a long GRB. Especially given the uncertain properties of the relativistic jets and the surrounding medium in each burst, this is not sufficient to begin to test whether the emission was simply not bright enough to observe with current facilities, if some (or all) pulses are obscured by the local environment, or whether these pulses simply do not occur. Long GRB durations are highly variable, and together with the broad redshift distribution, this means the timing of the radio observations are important. Yet, the triggered radio observations are all performed identically, and no long GRB rapidly observed at low frequencies thus far has had a secure redshift at the time of trigger. The dispersion measure through differing sightlines through the Milky Way, and especially host galaxy, will be different for each burst and a more complete study can be beneficial for understanding this better. This method has the potential to probe GRB jet composition, shock locations and environments, given a sufficiently large set of observations, such that the parameters going into the modelling can be disentangled.

    In this paper, we present an analysis of LOFAR HBA data of a further four long gamma-ray bursts, more than doubling the total number of LGRBs followed up: GRB\,200925B, GRB\,210104A, GRB\,240414A and GRB\,240418A, all bursts observed as part of a rapid response campaign. GRB\,210104A and GRB\,240414A include reported redshifts as part of other follow-up.
    
    In Section \ref{sec:observations}, we detail the \swift\ and LOFAR observations for each burst. Section \ref{sec:methods} describes the methods used to analyse and present the data. In Section \ref{sec:magneticmodel} we describe the magnetic wind model that predicts coherent radio emission to be emitted analogous to gamma-ray pulses. Section \ref{sec:lc_analysis} presents the light curves and results of the collected data in the context of the magnetic wind model - we discuss how we can constrain parameters due to non-significant detections of radio emission, and in one case assuming a low significance bump may be a real radio flare, and the limitations leading to non-observable radio flares. We discuss all the LGRBs from the LOFAR campaign as a whole in Section \ref{sec:lofarsample} alongside estimates of \eb in other literature. Finally, we conclude in Section \ref{sec:conclusion}.

\section{Targets}
    \label{sec:observations}
    
    Here we summarise the key information for the four long GRBs presented in this work. The burst durations are defined by the T90, the time period for 5 to 95 per cent of photons from the prompt phase to be detected.

    \begin{figure*}
        \includegraphics[width=\textwidth]{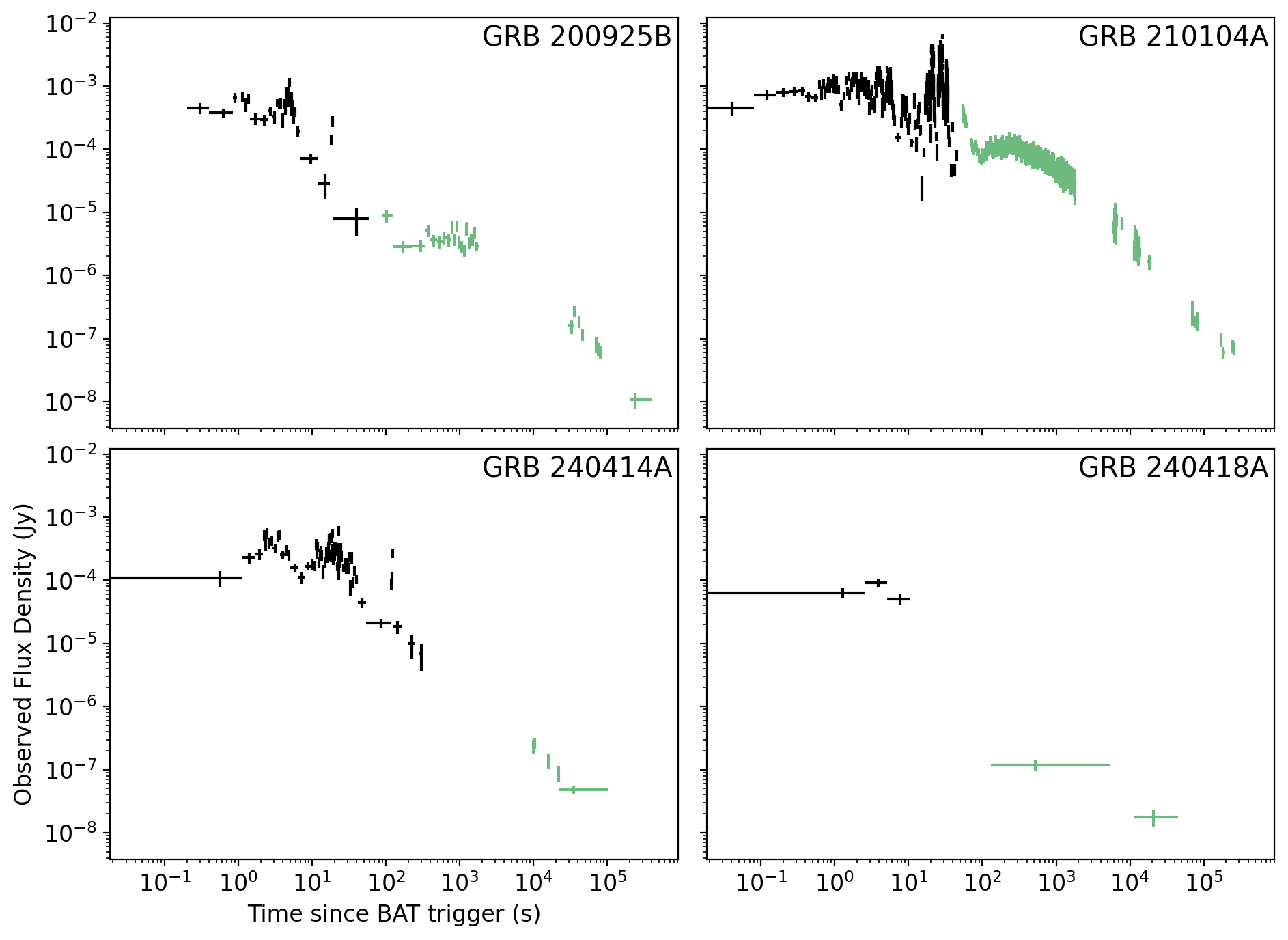}
        \caption{The observed flux density light curve for \swift-BAT (at 50\,keV, \textit{black}), and \swift-XRT (at 1\,keV, \textit{green}) for the four GRBs presented in this paper.}
        \label{fig:joint_lc}
    \end{figure*}

    \begin{figure*}
        \includegraphics[width=\textwidth]{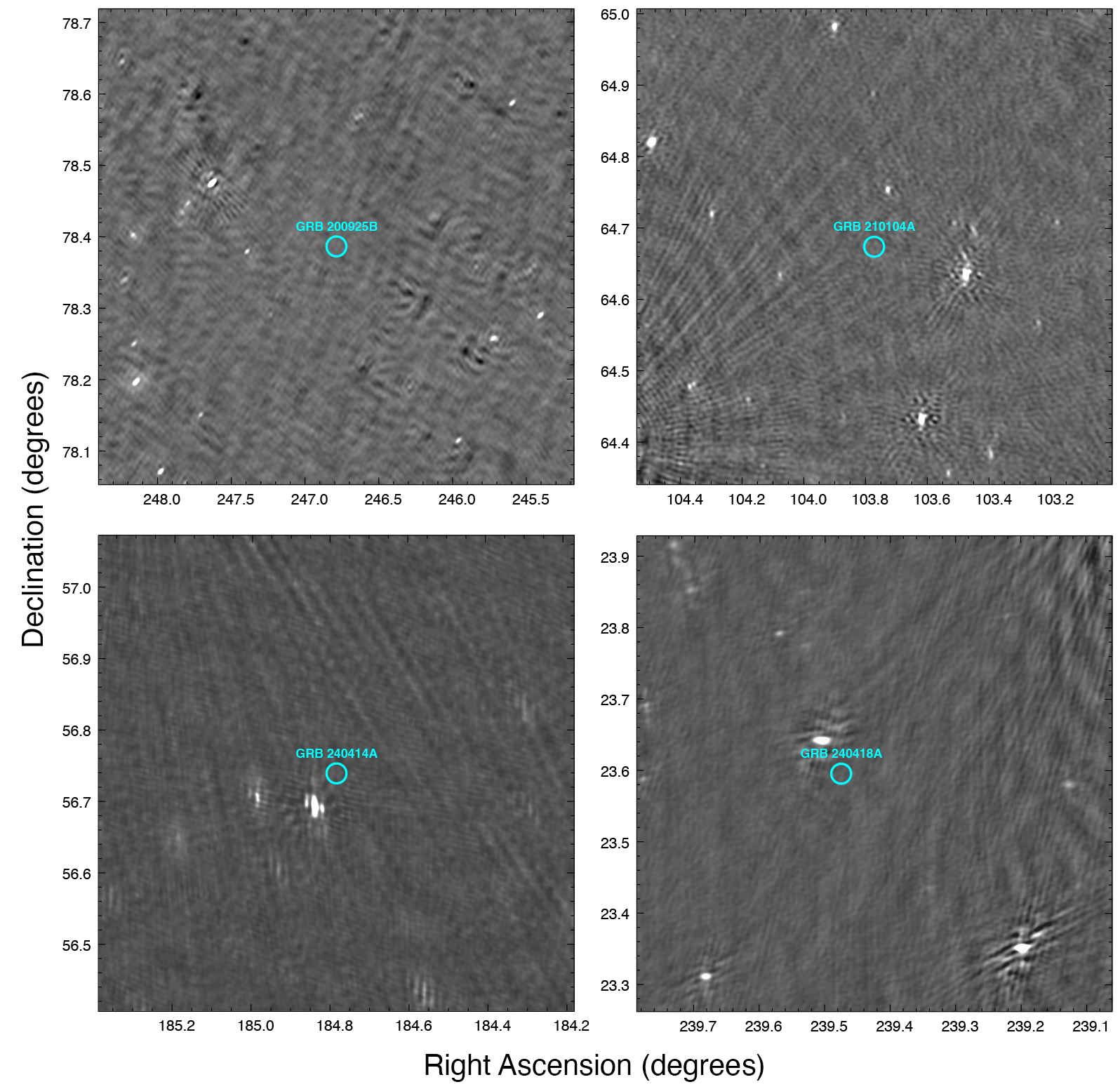}
        \caption{The deep LOFAR images spanning the full 2 hours of available data for each GRB dataset. From top-left clockwise: GRB\,200925B, GRB\,210104A, GRB\,240418A and GRB\,240414A. All images are $40\times40$\,arcmin. The blue circles are the position of the GRB, the size of the circles are 50 arcsecond and are only meant to guide the eye.}
        \label{fig:radioimages}
    \end{figure*}

        \subsection{GRB 200925B}
            \label{sec:obs_200925b}
    
            GRB\,200925B triggered the \swift\ Observatory on the 25\textsuperscript{th} September 2020 at 21:50:37 UT. The T90 in the 15--350\,keV range is reported as $18.25 \pm 0.97$\,s, designating this burst as an LGRB \citep{gcn_stamatikos2020}. The Burst Alert Telescope \citep[BAT,][]{barthelmy2005} was triggered by this burst and detected for $\sim 40$\,s. Simultaneously, the spacecraft was slewed and the X-ray Telescope instrument \citep[XRT,][]{burrows2005} on board began collecting data 78.4\,s after the initial trigger \citep{gcn_sbarufatti2020}.
    
            No redshift was obtained for this burst. An optical afterglow was found in I band \citep{gcn_deugartepostigo2020} placing an upper limit of $z \lesssim 5.5$.
    
            LOFAR's rapid response mode was automatically triggered, and with the GRB position being immediately observable, LOFAR HBA began observing the target location at 21:55:04 UT, 267 seconds after the initial BAT trigger, collecting 2 hours of imaging data.

        \subsection{GRB 210104A}
            \label{sec:obs_210104a}

            \swift-BAT was triggered at 11:26:59 UT on 4\textsuperscript{th} January 2021, collecting $\sim40$\,s of gamma-ray data. The prompt emission shows a highly variable structure with a number of distinctive peaks. The reported T90 is $32.06 \pm 0.49$\,s \citep{gcn_palmer2021}. \swift-XRT began observations 69 seconds after the initial trigger \citep{gcn_kennea2021}.
            
            LOFAR automatically triggered at 11:31:28 UT and began collecting 2 hours of imaging data, 269 seconds after the BAT trigger.

            For this burst there is a tentative redshift of $z=0.46$ reported through spectroscopic observations of absorption features \citep{zhang2022a}.

        \subsection{GRB 240414A}
            \label{sec:obs_240414a}

            GRB\,240414A occurred at 02:20:36 UT on the 14\textsuperscript{th} April 2024 after triggering \swift-BAT. The prompt emissions show a similar structure to that of GRB\,200925B, characterised by several pulses, and a final short pulse during the rapid decay phase before giving way to afterglow emissions. The T90 as reported by BAT was $88.28\pm 50.41$\,s \citep{gcn_markwardt2024}. BAT collected about 300 seconds of data, and the spacecraft slewed and an XRT position was promptly available, but XRT data are only available from 10\,k seconds \citep{gcn_dichiara2024}.
            
            A redshift of $z=1.833$ is reported for this burst as the result of optical observations of a number of absorption features \citep{gcn_adami2024,gcn_deugartepostigo2024}, observed by 
            the Multi-purpose InSTRument for Astronomy at Low-resolution \citep[MISTRAL,][]{schmitt2024} and GTC Optical System for Imaging and low Resolution Integrated Spectroscopy \citep[GTC/OSIRIS+,][]{cepa1998}.

            LOFAR triggered at 02:26:49 UT and began collecting 2 hours of follow-up imaging data, 373 seconds after the initial BAT trigger.

        \subsection{GRB 240418A}
            \label{sec:obs_240418a}

            \swift-BAT triggered on the 18\textsuperscript{th} April 2024 at 20:24:08 UT with a reported T90 of $12.00 \pm 3.61$\,s \citep{gcn_palmer2024}. \swift-XRT began observing 122 seconds after the BAT trigger, revealing a featureless, decaying afterglow in 6.1\,ks of X-ray data \citep{gcn_tohuvavohu2024}. Several optical facilities were unable to detect an optical counterpart with limiting magnitudes of $r>22$, $z>20.4$ and $J>19$, but NIRES identified an afterglow candidate in the K' filter with magnitude $20.9\pm0.1$ \citep{gcn_karambelkar2024}.
            
            Imaging with LOFAR's rapid response mode was triggered 373 seconds after the initial trigger at 20:30:21 UT.

\section{Methods}
\label{sec:methods}

    Here we detail the methods used to present and analyse the gamma-ray, X-ray and radio data.

    \subsection{\swift\ data}
    \label{sec:methods_swift}

        In order to produce predictions for the flux density of radio flares emitted analogous to prompt emission pulses, described later in Section \ref{sec:magneticmodel}, the light curves and flares need to be modelled to calculate the fluence from each flare component. 

        We obtained the flux density data for BAT and XRT (at 50\,keV and 1\,keV, respectively) from the online \swift\ archive\footnote{\url{https://www.swift.ac.uk}} at the UK Swift Science Data Centre \citep[UKSSDC, see][]{evans2007,evans2009}. XRT data uses all available observation modes where provided. The combined light curves for each burst are displayed in \autoref{fig:joint_lc}.

        Selected light curves are fitted with our code that models the flares and afterglows of GRBs, described in \citet{hennessy2023}. To summarise, the code first identifies flares by requiring that at least two out of three consecutive data points rise at least twice the first point's 90 per cent confidence interval. Start, peak and end times are found through local minima, maxima and calculating the changing gradient, and excluded. The remaining data represents the afterglow and is fitted with a broken power law model of up to 5 breaks. Finally, the flares are modelled as FRED peaks, described by:
        \begin{equation}
            \label{eqn:fred}
            f(t) = A \times \sqrt{\exp\left(2\times\frac{R}{D}\right)} \times \exp\left(-\frac{R}{t-t_{s}} - \frac{t-t_{s}}{D}\right)\,,
        \end{equation}
        where $A$ corresponds to amplitude, $R$ and $D$ are variables determining the rise and decay steepness and $t_{s}$ is the start time of the modelled flare. FRED flares represent a fast energy release followed by some slower diffusive process and are associated with several transient phenomena, specifically, they model well the prompt emission pulses in many GRBs \citep{norris1996,ma2024}.

        Since the data collected are in units of observed flux density in this case, to fit BAT and XRT data simultaneously, we must convert to units of flux in each instruments' native frequency in order to integrate and measure a fluence. Spectral evolution data from the power law fits to the spectra are made available from the \swift\ Burst Analyser \citep{evans2010}, allowing us to convert from flux density at 50\,keV to flux in the 15-50\,keV energy range for the BAT data. The flares we are modelling are only present in the BAT data for this study, but we model and plot the \swift-XRT for completeness - available from the \swift\ light curve repository \citep{evans2007,evans2009}.

    \subsection{LOFAR data}
        \label{sec:methods_lofar}

        \begin{table*}
            \caption{A summary of the LOFAR observations.}
            \label{tab:lofarobs}
            \begin{tabular}{lrrrrr}
            \hline
            GRB & Project Code & Observation ID & Calibrator & Start Time & Duration (s) \\ \hline
            200925B & LC14\_004 & 796116 & 3C 295 & 21:55:04 & 7199 \\
            210104A & LC15\_013 & 815080 & 3C 147 & 11:31:28 & 7199 \\
            240414A & LC20\_021 & 2039208 & 3C 295 & 02:26:49 & 7200 \\
            240418A & LC20\_021 & 2039344 & 3C 286 & 20:30:21 & 7200 \\\hline
            \end{tabular}
        \end{table*}

        LOFAR rapid response observations were initiated automatically using a custom-built script to parse \swift\ GCN Notices (using {\sc voevent-parse} tools, \citealp{staley2014}\footnote{\url{https://voevent-parse.readthedocs.io/en/stable/}}), select appropriate GRBs and communicate with LOFAR. Trigger criteria demanded that the GRB be observable to LOFAR within ten minutes of the Swift BAT trigger at an elevation of $\ge20^{\circ}$ for at least 30 minutes, with an optimal calibrator source simultaneously available. We also required a BAT rate trigger significance of $\ge10\sigma$ (the photon count rate significance over background noise) or rate trigger integration timescale of $\le2$\,s (the event duration likely represents a short gamma-ray burst), and a prompt \swift-S/C Will$\_$Slew notice\footnote{A GCN notice that \swift\ will slew to the GRB position and continue to observe the target.} to ensure X-ray Telescope observations and hence a refined localisation on the order of arcseconds.

        All observations were made in the 120--168\,MHz frequency range using the LOFAR HBA. The data were averaged with a 1\,s integration time in 244 sub-bands of 195.3\,kHz. The Dutch array was used, composed of 23 core stations and 11 remote stations, yielding a maximum baseline of 121\,km \citep{vanhaarlem2013}. After the target observations, a 10 minute calibration observation was taken.

        The data were averaged to 8 second time resolution then calibrated and imaged using the LOFAR Initial Calibration ({\sc LINC}, version 4.0)\footnote{\url{https://linc.readthedocs.io/en/latest/}} pipeline, used to correct for instrumental effects in LOFAR observations. Standard software and methods were used, as described in \citet{vanweeren2016,williams2016,degasperin2019}. Part of the process includes {\sc AOFlagger} \citet[version 3.3,][]{offringa2010,offringa2012} to allow statistical flagging and removal of data affected by radio interference. We used a calibrator observation (see \autoref{tab:lofarobs}) to derive gain solutions, which are transferred to the target observation. A phase-only calibration was applied to the target data using a sky model from TGSS ADR1\footnote{Tata Institute of Fundamental Research Giant Metrewave Radio Telescope Alternative Data Release} \citep{intema2017}. We decided that further direction dependent calibration was not required in any case, as the GRBs were at the centre of the field in each observation.
        
        The calibrated files were imaged with {\sc WSClean} \citep[version 3.1.1,][]{offringa2014}\footnote{\url{https://wsclean.readthedocs.io/en/latest/}}, a wide-field interferometric imager. Standard imaging parameters were used: Briggs weight -0.5 and auto-thresholding to 100,000 iterations. A pixel scale of 1 arcsecond was used with a $2048 \times 2048$ pixel image size.

        A deep image was created for each LOFAR observation, showing the GRB and surrounding region, shown in \autoref{fig:radioimages}, to look for any persistent emission at the GRB position, such as from the host galaxy. Time sliced images using the full frequency range were also created to produce radio light curves, at 60 and 180 second intervals, using the same imaging parameters. These time lengths were selected based on a dispersed flare length of $\sim 300-400$ seconds \citep{starling2020,hennessy2023}. A choice of bin width equal to the radio flare width would result in only one elevated point per flare, which would hamper positive identification, while choosing a width much less than half the flare width runs the risk of low signal-to-noise per bin also hampering identification. Consequently, a 60 second time slice may allow us to observe the rise and fall structure of a radio flare.
    
        The LOFAR Transients Pipeline \citep[{\sc TraP}, version 6.0,][]{ LINC}\footnote{\url{https://docs.transientskp.org}} is a pipeline that looks for transient or variable sources within astronomical data, primarily designed for LOFAR data. We used {\sc TraP} to obtain flux measurements in each image using a force fit extraction at the GRB position. {\sc TraP} also outputs a background root mean square (rms) noise in each image, calculated using the inner 1/8th of the image. For time sliced images, source extraction is undertaken for each slice using default settings and restoring beam, allowing us to create a radio light curve at the location of the GRB.
                
        Where there was a flux density enhancement found, we created images across 6 frequency channels using the 60 second interval data to create individual light curves at each channel frequency, as a means of looking for a dispersed signal.

\section{Magnetic wind model}
    \label{sec:magneticmodel}

    The \citet{usov2000} model describes a relativistic, strongly magnetised wind that interacts with the negligibly magnetised intermediate media surrounding the burst, creating a surface current at this boundary with a very strong electric field. This accelerates electrons in the plasma to relativistic speeds, resulting in the synchrotron spectrum observed at X-ray and gamma-ray wavelengths. Additionally, the model also predicts low-frequency electromagnetic waves to be produced at the same time due to a varying electric field from the non-stationarity of the interaction. We should expect to see radio emission emitted during the prompt emission phase similar to how we see pulses in gamma-rays.
    
    \citet{starling2020} applied this model to existing data, predicting that 44\% of X-ray flares in a sample of \swift\ GRBs with redshifts should have a detectable prompt radio component at 144 MHz with LOFAR HBA, assuming that X-ray flares arise from the same internal mechanisms as gamma-ray pulses. We follow the method of a previous analysis of a single GRB studied through this LOFAR follow-up campaign, described in \citet{hennessy2023}. 
    
    The emissions from the source are dispersed due to the interactions with intermediate matter along the line-of-sight to the GRB, and hence the radio emission is delayed relative to the gamma-rays. This delay $\tau(\nu)$ is estimated as:
    \begin{equation}
        \label{eqn:dispersion}
        \tau(\nu) \sim \frac{\textrm{DM}}{241\nu^2}\,\textrm{s}\,,
    \end{equation}
    where DM is the dispersion measure (units of pc\,cm$^{-3}$) and $\nu$ the observing frequency in GHz \citep{taylor1993}.

    DM is expected to scale with the number of free electrons along the line-of-sight. Therefore, a larger DM is expected for sources at a larger distance, or for sources with particularly dense host environments. In this paper, we follow \citet{lorimer2007} and other similar LOFAR rapid response studies of GRBs and take DM as $\sim1200z$\,pc\,cm$^{-3}$ \citep{rowlinson2019, hennessy2023}.
        
    The predicted flux density for a radio pulse, in the dispersion limited case, based on the fluence in gamma-ray pulses is given (by \citealp{usov2000}) as:
    \begin{equation}
        \label{eqn:radioflux}
        F_{v} = \frac{\delta(\beta-1)}{2\Delta \nu \tau(\nu)}\left(\frac{\nu}{\nu_{\textrm{max}}}\right)^{1-\beta} \frac{\Phi_{\gamma}}{10^{-23}}\,\textrm{Jy}
    \end{equation}
    where $\delta \sim 0.1\epsilon_{B}$ is the ratio of bolometric radio fluence (120-168\,MHz) to bolometric $\gamma$-ray fluence (15-50\,keV); $\beta$ is the spectrum power law index in the high frequency radio tail ($\gtrsim 0.01$\,MHz); $\Delta\nu$ is the observing bandwidth (48\,MHz); $\nu_\textrm{max}$ is the peak frequency of radio emission \citep[see equation 1 in][]{hennessy2023} and $\Phi_{\gamma}$ is the bolometric $\gamma$-ray fluence in erg\,cm$^{-2}$. The emission occurs over a range of frequencies, and hence will experience broadening to due the effects of dispersion delay varying based on frequency. The form of \autoref{eqn:radioflux} accounts for this broadening in the dispersion limited regime. For all flares in this study, we are in this regime, where the dispersion broadening dominates over the intrinsic flare length, in this case assumed to be the same length as the gamma-ray pulse.
    
    As a result, we are able to analyse the LOFAR radio datasets, and test predictions of this model for bursts with prompt gamma-ray pulses for which the redshift is known.

\section{Results of the search for coherent radio emission}
    \label{sec:lc_analysis}

    \begin{table*}
        \caption{Summary of the light curves fitted with our code for the two GRBs with a redshift, \autoref{fig:grb200925b_laffradio} and \autoref{fig:grb240414a_laffradio}. The afterglow is fitted with a broken power law, and the flares fitted with a FRED curve \autoref{eqn:fred}, as described in Section \ref{sec:methods_swift}$^{*}$.}
        \label{tab:flareparams}
        \begin{tabular}{lcccccrrrc}
            \hline
            GRB & Flare \# & Temporal index & \multicolumn{3}{c}{Flare parameters$^{*}$} & \multicolumn{3}{c}{Flare start/peak/end} & Flare $\gamma$-ray fluence (15-50\,kev) \\
             &  & under flare & $t_{s}$ & R & D & \multicolumn{3}{c}{s} & erg\,cm$^{-2}$ \\ \hline
            200925B & 1 & 0.05 & 3.42 & 8.53 & 0.19 & 2.24 & 4.94 & 6.34 & $6.33 \times 10^{-8}$ \\
             & 2 & 2.44 & 18.16 & 0.002 & 5.13 & 14.80 & 19.06 & 64.99 & $1.51 \times 10^{-7}$ \\
            240414A & 1 & 0.05 & -1.17 & 25.53 & 0.60 & 0.56 & 2.46 & 7.30 & $9.63 \times 10^{-8}$ \\
             & 2 & 0.05 & -0.41 & 113.26 & 2.66 & 7.30 & 22.92 & 30.24 & $4.00 \times 10^{-7}$ \\
             & 3 & 1.42 & 111.93 & 92.36 & 3.00 & 85.48 & 122.10 & 222.28 & $5.15 \times 10^{-7}$ \\ \hline 
        \end{tabular}
    \end{table*}

    \begin{figure*}
        \includegraphics[width=\textwidth]{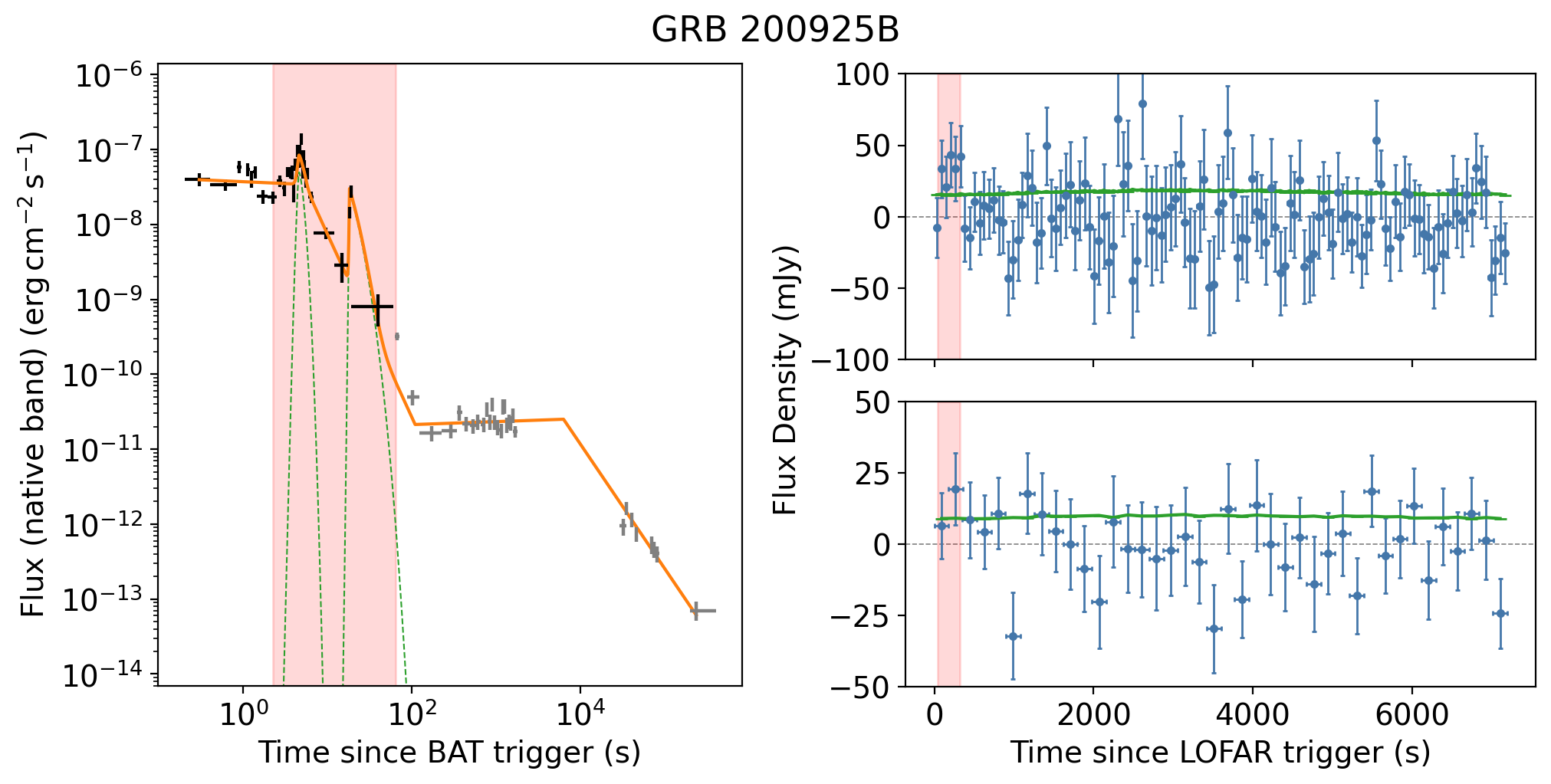}
        \caption{(Left) the flux light curve of GRB\,200925B for BAT data (15-50\,keV, \textit{black}) and XRT data (0.3-10\,keV, \textit{grey}). Overlaid is the result of the GRB modelling curve described in Section \ref{sec:lc_analysis} (and in further details in \citealp{hennessy2023}) - flare (\textit{green}) components are shown and total model fit (\textit{orange}) are shown. \autoref{tab:flareparams} contains a summary of the fit. Three power law breaks and two flares are found for this burst. (Right) the observed 144\,MHz radio flux density at the position of GRB\,200925B as a function of time. Top panel shows the 60 second time sliced images, and bottom shows the 180 second time sliced images. The rms noise in each image is shown in green. The red shaded region shows the predicted timing and duration of a radio flare overlapping a $2\sigma$ bump, which corresponds to the red shaded region in the \swift\ light curve, for $z\sim1.8$.}
        \label{fig:grb200925b_laffradio}
    \end{figure*}

    \begin{figure}
        \includegraphics[width=\linewidth]{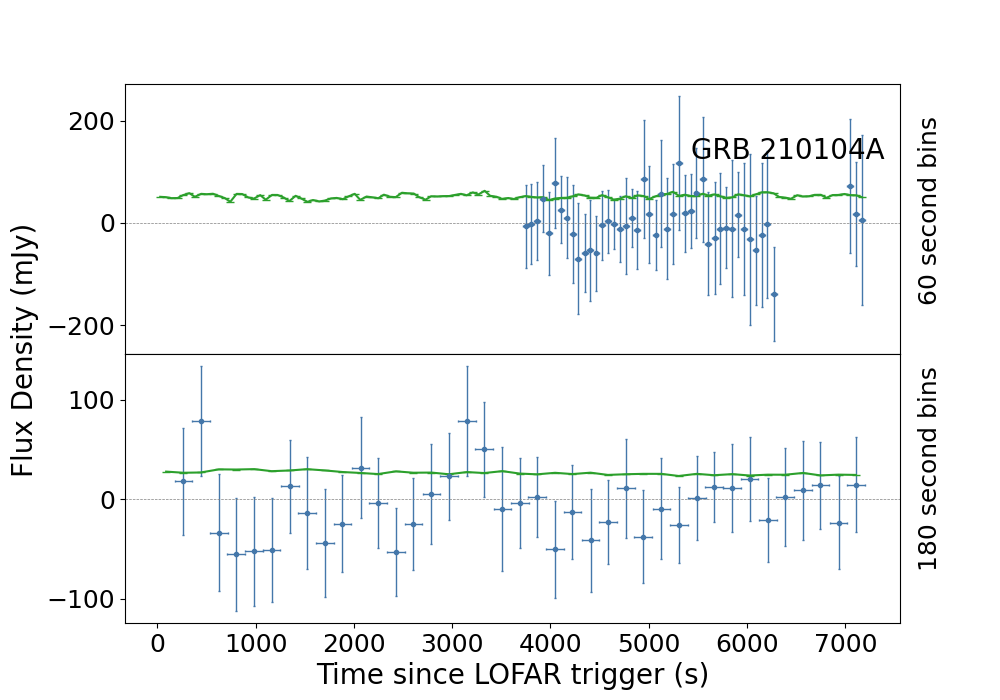}
        \caption{The observed 144\,MHz flux density at the position of GRB\,210104A as a function of time, for 60 second and 180 second time slices. The green line shows the $1\sigma$ rms noise measured in each image.}
        \label{fig:radio_210104a}
    \end{figure}

    \begin{figure*}
        \includegraphics[width=\textwidth]{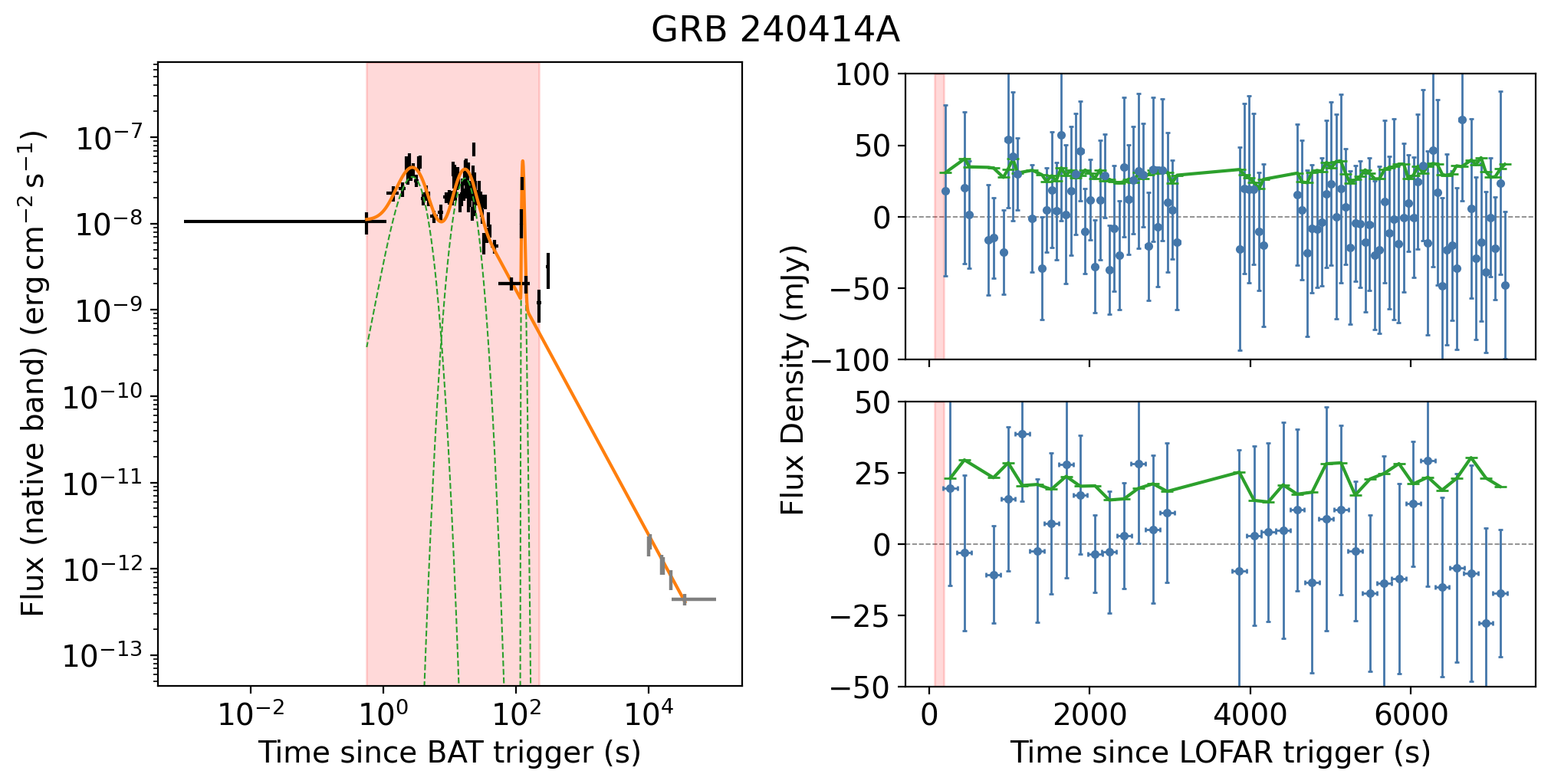}
        \caption{(Left) the flux light curve of GRB\,240414A for BAT data (15-50\,keV, \textit{black}) and XRT data (0.3-10\,keV, \textit{grey}). Overlaid is the result of the GRB modelling described in Section \ref{sec:lc_analysis} (and in further details in \citealp{hennessy2023}) - flare (\textit{green}) components are shown as well as total model fit (\textit{orange}). One power law break and three flares are found for this burst. (Right) the observed 144\,MHz radio flux density at the position of GRB\,240414A as a function of time. Top shows the 60 second time sliced images, and bottom shows the 180 second time sliced images. The rms noise in each image is shown in green. The red shaded region shows the predicted timing and duration of a radio flare corresponding to the reported redshift of $1.833$. Some data points are missing due to poor quality data.}
        \label{fig:grb240414a_laffradio}
    \end{figure*}

    \begin{figure}
        \includegraphics[width=\linewidth]{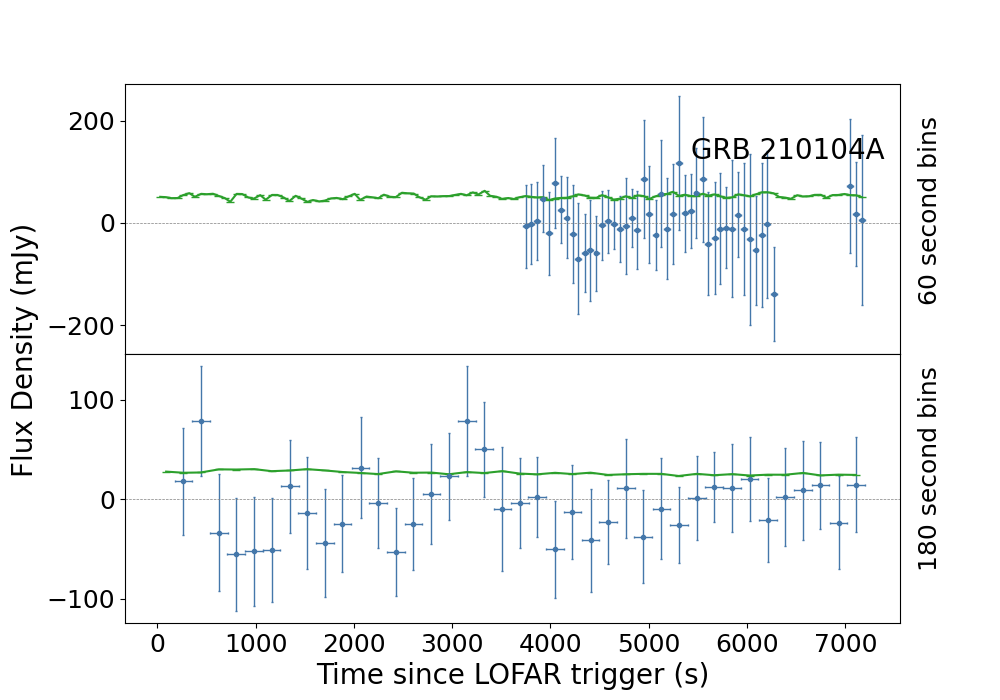}
        \caption{The observed 144\,MHz flux density at the position of GRB\,240418A as a function of time for 180 second time slices. The green line shows the $1\sigma$ rms noise measured in each image.}
        \label{fig:radio_240418a}
    \end{figure}

    \subsection{GRB 200925B}

        The results for GRB\,200925B are shown in \autoref{fig:grb200925b_laffradio}, where we fit the BAT-XRT light curve and present the 144\,MHz radio flux light curves. The high energy light curve is well modelled with three breaks in the continuum, showing a fairly standard 'canonical' GRB, as described in \citet{evans2009}; and two flares seen in the BAT data during and just after the prompt emission phase, peaking at 4.9\,s and 19.1\,s. The measured 15--50\,keV fluences for the two flares are $6.33\times10^{-8}$ and $1.51\times10^{-7}$ erg\,cm$^{-2}$, respectively. See \autoref{tab:flareparams} for a summary of the modelled flares.

        In both 60 second and 180 second binned light curves, we observe no significant radio emission. We note a small $2\sigma$ (twice the rms) bump in both light curves at $\sim120$\,s, seen in one 120\,s image and five consecutive 60\,s images. We searched for a dispersed signal over the time of this bump by looking at each frequency channel individually. Consecutively delayed peaks are expected, with the lower frequencies arriving later, due to dispersion.

        We expect the two flares seen in the high-energy regime to produce a corresponding signal at radio frequencies in line with the model discussed in Section \ref{sec:magneticmodel}. We require a minimum value of $z \gtrsim 1.05$, resulting in a dispersion delay of 252\,s, for both analogous radio peaks to be sufficiently delayed into the LOFAR observation window, given the 267\,s delay for observations to start compared to the BAT trigger. Alternatively, $z \gtrsim 1.10$ produces conditions for just the latter, more energetic peak to become observable in the data, with a dispersion delay of 265\,s in this case. However, $z=1.8$ produces a dispersion delay of 428\,s which is sufficient for the predicted timing of the flare to fall at the time of the small bump seen in one 180\,s and several consecutive 60\,s bins. Looking at the timing of arrival of maximum flux in each frequency channel, a dispersed signal is not evidenced.

        The observed pulse width will be longer than the intrinsic pulse width as a result of dispersion - the pulse occupies a finite frequency range, where each part is dispersed by a slightly different amount. By modelling the temporal shape of the intrinsic pulse as a delta function we estimate the observed duration as 285.4\,s at $z=1.8$ \citep[equation 3,][]{hennessy2023}. The real pulse will have some finite length, so we expect this to offer a lower limit on the total duration of observed radio emission, but expect the peak emission to occur in this time frame. Given the width and separation of the two gamma-ray pulses seen in the \swift\ light curve (\autoref{tab:flareparams}), we expect the radio pulses to be almost totally overlapping.

        Assuming a non-detection of radio emission in these images, we obtain an output rms value of 17.1 and 9.6\,mJy to derive a $3\sigma$ ($3\times$ rms) upper limit on any flaring activity of 51.3 and 28.8\,mJy, for integration lengths of 60\,s and 180\,s, respectively.

        Subsequently, in Section \ref{sec:discussion_200925b}, we discuss the implication of a non-detection on the input parameters of our model. We also consider the case and implications of a radio signal being observed, in line with the observed, low significance bump in the radio light curve.

    \subsection{GRB 210104A}

        The high energy afterglow modelling finds 4 power law breaks and 16 flares in the energetic prompt phase. However, these flares are not accessible due to the combination of the reported redshift of $z=0.46$ producing a dispersion delay of 110.5 seconds and LOFAR's 269 second response time - in this case the predicted timing of the radio flares would be before observations began. Our LOFAR observation begins only during the plateau phase of the afterglow. We note, however, there is a possibility the redshift is higher, up to a limit of about $z=2.2$, due to no significant host being found and based on the lack of clear Lyman-alpha line in the optical spectrum.
        
        The 144\,MHz radio flux light curve is shown in \autoref{fig:radio_210104a}. In both 60 second and 180 second binned images, we do not detect any significant radio signal for this burst, and we obtain an output rms value of 24.5 and 15.1\,mJy to derive $3\sigma$ upper limits of 73.5 and 45.3\,mJy, for integration lengths of 60\,s and 180\,s, respectively.

    \subsection{GRB 240414A}

        GRB\,240414A is found to have one power law break and three flares, in the high energy data when fitted with our light curve modelling code. The prompt phase shows several energetic pulses. The flares peak at 2.46, 22.92 and 122.10 seconds with integrated 15--50\,keV fluences of $9.63\times 10^{-8}$, $4.00\times 10^{-7}$ and $5.15\times 10^{-7}$ erg\,cm$^{-2}$, respectively. The final few BAT data points may indicate the existence of another flare, but there is insufficient data to attempt to model this.

        The reported redshift $z=1.833$ produces a dispersion delay of $\sim440$ seconds. LOFAR's response time of 373 seconds means that we should expect to see the equivalent radio emission in the first few bins of our radio light curve - though the {\sc TraP} source extractor is unable to produce a data point for some time slices due a large percentage of the data being flagged, representing poor quality data, usually due to interference from atmospheric conditions.

        The fit of the \swift-BAT and XRT light curve data, and radio light curve are displayed in \autoref{tab:flareparams} and \autoref{fig:grb240414a_laffradio}. In 60 second and 180 second binned sets of images, we observe no significant radio emission. The quality of some parts of this observation were lower, leading to heightened flagging in the {\sc LINC} pipeline. The first 180 second (and first three 60 second) bins were unable to be imaged, cutting off some data during the predicted radio timing. We should still expect to see most of the emission from the latest, most energetic flare. For a higher fraction of energy in the magnetic field, $\epsilon_{B} = 10^{-3}$, the radio flux density is predicted to be 281\,mJy just for the third flare, which is not observed. However, a more conservative estimate of $\epsilon_{B} = 10^{-4}$ predicts 14\,mJy, in which case the emission would be obscured in the noise consistent with our findings.

        Regardless of the true $\epsilon_{B}$ value, we can place an upper limit on any radio emission at this time - an image rms of 30.7\,mJy produces a $3\sigma$ limit of 92.2\,mJy for 60 second bins, and a rms of 21.8\,mJy produces a $3\sigma$ limit of 65.4\,mJy on any 180 second emission.

    \subsection{GRB 240418A}
    
        The \swift\ BAT-XRT data for this burst showed very little structure and there is no reported redshift meaning we are unable to directly test the magnetic model with this data set. In this case, the data quality was not sufficient to image on 60\,s timescales.

        We have nevertheless imaged the data anyway to test for any radio emission and shown this in \autoref{fig:radio_240418a}. We report no significant emission in our images, but are able to derive a $3\sigma$ upper limit of 79.5\,mJy for an integrated length of 60 seconds.

   \subsection{Low significance enhancement in GRB 200925B}
        \label{sec:discussion_200925b}
    
        In GRB\,200925B, we note a $\sim2\sigma$ increase in flux density at a time that would be analogous to the gamma-ray pulses emitted in the prompt emission for $z=1.8$. The predicted duration of each flare at this redshift is $\sim 258$ seconds. The proximity of the flares in the \swift-BAT data means we should expect the radio flares to be observed overlapping and hence expect a slightly longer duration, and a summed 144\,MHz flux density from both flares. This aligns well with the $\sim 300$ second duration of the enhancement.

        We use the gamma-ray fluences of our GRBs to generate radio flux density predictions. Following \citet{usov2000} (also used in \citet{starling2020}) we take $\delta\sim0.1\epsilon_{B}$ and $\beta\sim 1.6$. If we follow \cite{katz1997} and take $\epsilon_{B} = 10^{-3}$, using \autoref{eqn:radioflux}, we obtain 144\,MHz flux density predictions of 32.0 and 76.3\,mJy for the first and second flare, respectively. While the prediction for the first flare would remain hidden within the noise of the data, the second should be observable, and certainly the combined radio flare should be visible. However, this is over-predicting the enhancement observed in the radio light curve. 
        
        The duration and timing of the low significance radio flare would be consistent with coming from the prompt pulses of GRB\,200925B if it lies at $z=1.8$. In the following section, we look at how lowering \eb can better reproduce our observations.

    \subsection{Constraining the fraction of energy in the magnetic field}

        \begin{figure}
            \includegraphics[width=\columnwidth]{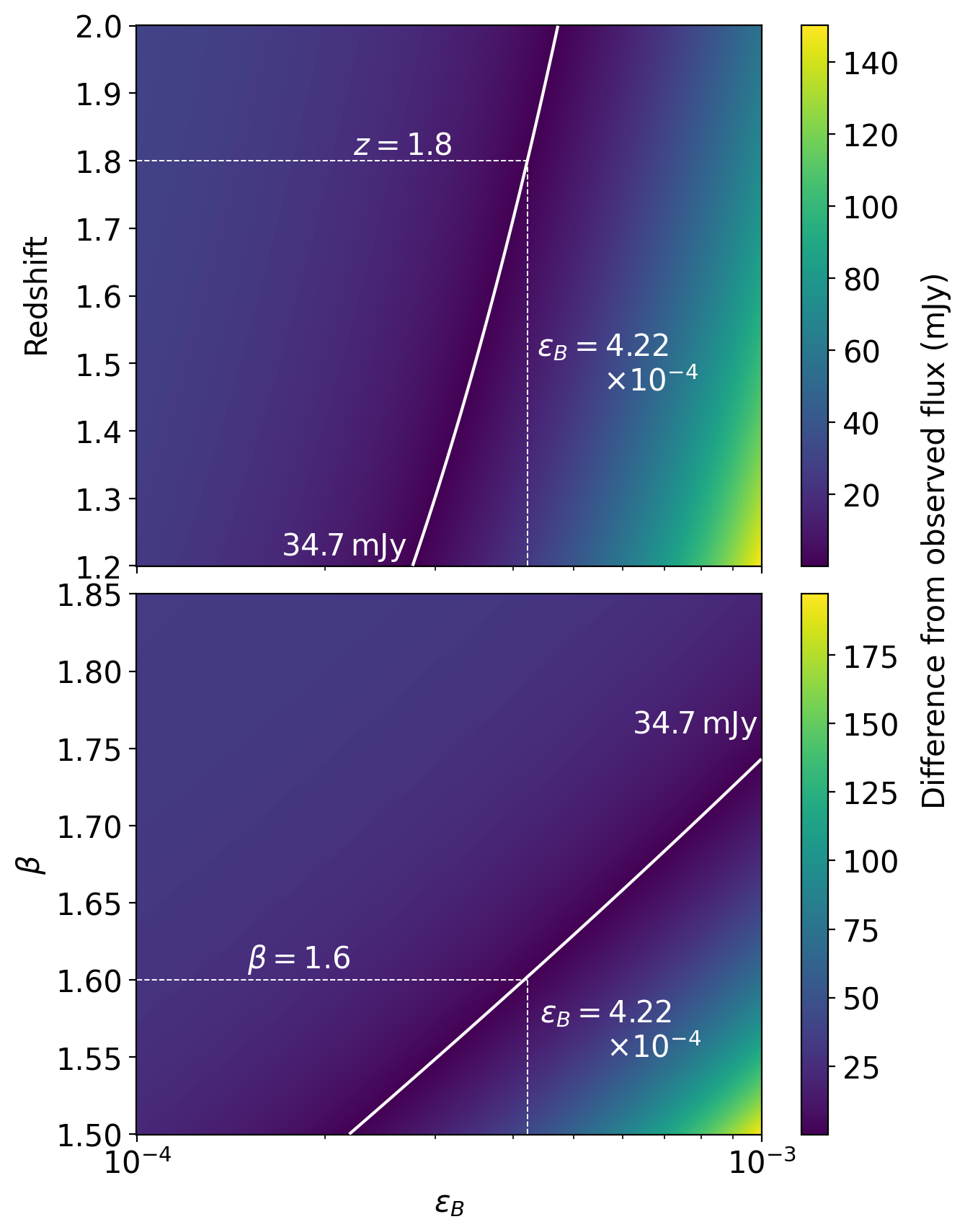}
            \caption{The parameter space for $\epsilon_{B}$ against redshift for constant $\beta=1.6$ (top) and $\epsilon_{B}$ against $\beta$ for constant $z=1.8$ (bottom) for GRB\,200925B. The solid white represents the parameter set that produces the 'target' value of 34.7\,mJy, obtained as the average flux density across the small bump in the radio light curve. Dashed white line represents the redshift that provides the correct timing constraint for this bump, and correspondingly the required $\epsilon_{B}$ given this value. The colour scale represents how far off from the target value that parameter set produces.}
            \label{fig:parameters_200925b}
        \end{figure}

        \begin{figure}
            \includegraphics[width=\columnwidth]{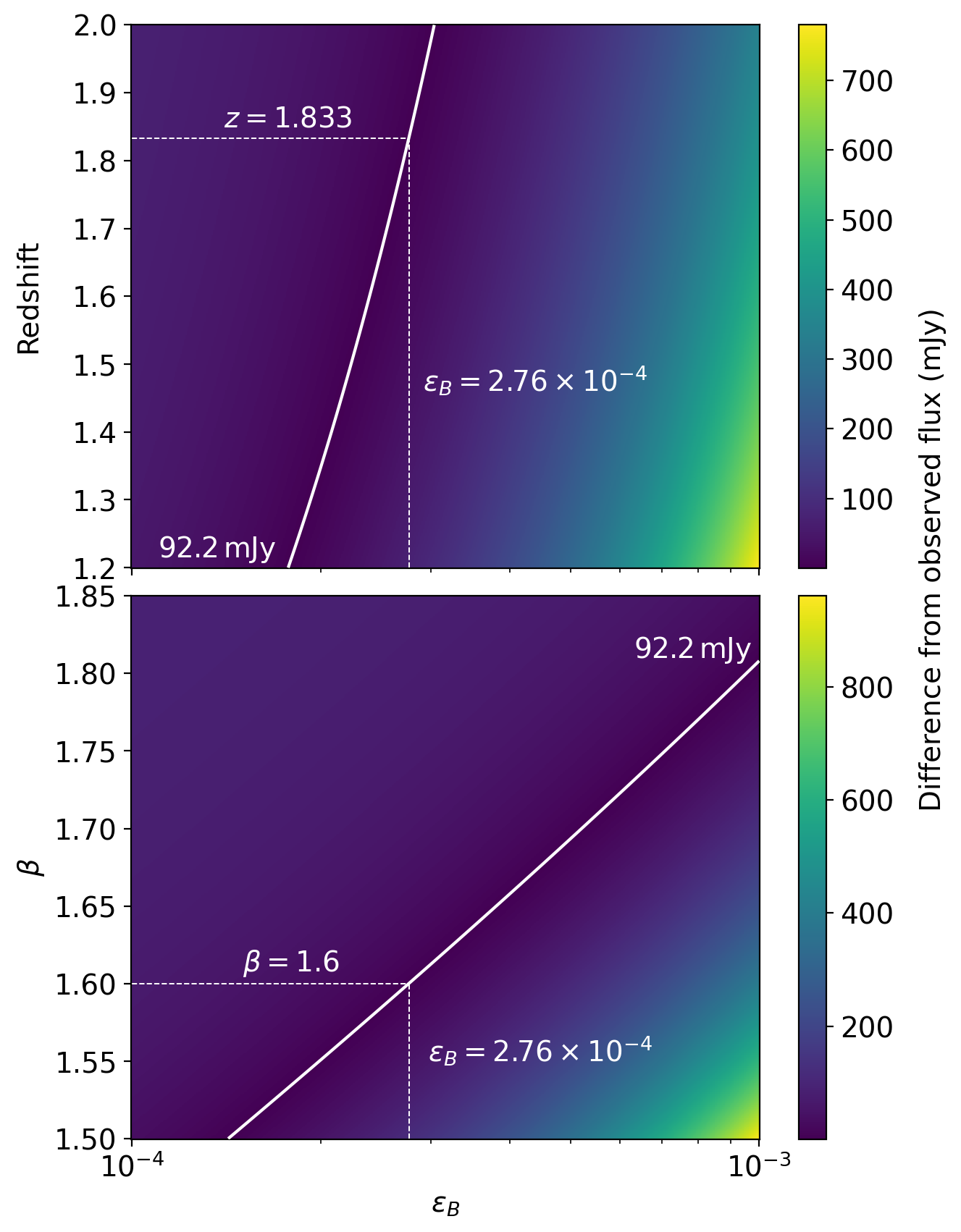}
            \caption{The parameter space for $\epsilon_{B}$ against redshift for constant $\beta=1.6$ (top) and $\epsilon_{B}$ against $\beta$ for constant $z=1.833$ (bottom) or GRB\,240414A. The solid white represents the parameter set that produces the 'target' value of 92.2\,mJy, obtained as $3\sigma$ image rms across the radio light curve. Dashed white line represents the redshift that provides the correct timing constraint for this bump, and correspondingly the required $\epsilon_{B}$ given this value. The colour scale represents how far off from the target value that parameter set produces.}
            \label{fig:parameters_240414a}
        \end{figure}

       In line with previous prompt studies, our results favour a lower fraction of energy in the magnetic field. LOFAR has the sensitivity to probe to this level, while challenges still remain in constraining microphysical parameters including $\epsilon_{B}$.
        
        Given the presence of pulses seen in the \swift-BAT light curve, the LOFAR trigger time and a redshift, we can produce constraints on \eb for both cases of GRB\,200925B and for GRB\,240414A by varying this parameter within the magnetic reconnection model prediction to match our observations.
        
        If we consider the flux density enhancement in the low-frequency light curve of GRB\,200925B to be real, we can produce radio flux predictions over a parameter space of ranging redshift, $\beta$ and $\epsilon_{B}$, shown in \autoref{fig:parameters_200925b}, by keeping the third parameter constant. The high frequency radio tail spectral index is kept constant at $\beta=1.6$, following \citet[][and references therein]{starling2020}. Redshift is kept constant at $z=1.8$, inferred from the delay between gamma-ray peak time and the start of the radio enhancement, testing the case in which the small flux density rise is a real radio flare. We can look for the parameter set that produces a radio flux density prediction equal to the enhancement seen in the radio light curve - any real radio emission emitted coincident with the gamma-ray pulse should be equal to this observed level, or less, which represents a radio flare hidden within the noise of the data. Any greater flux density prediction should be seen in the data, hence, we can use this to produce upper limits on parameters. The white line represents this target flux density value, while the colour scale represents how far off this value other parameter sets predict. The GRB target position in the images across the enhancement have an average value of $34.7$\,mJy. At a value of $\epsilon_{B} = 4.2 \times 10^{-4}$, we satisfy the conditions to produce this observed flux density of two overlapping flares, emitted analogous to the two prompt pulses seen in the BAT data, for a redshift of $1.8$. In the case of a complete non-detection, we can use our $3\sigma$ upper limit of 51.3\,mJy from the rms noise of our 60 second time sliced imaging to give an upper limit of $\epsilon_{B} < 5.6\times10^{-4}$ on any coherent radio flare activity at the $3\sigma$ significance level at $z=1.8$. We note that there is uncertainty in our redshift value due to a lack of a reported redshift, spectroscopic or otherwise, though this value sits reasonably in the distribution of range of \swift\ GRB redshifts \citep{evans2009,lan2021}.

        For GRB\,240414A, the quality of the data means the GRB position is not sampled across all the images, in particular some images during the expected arrival time of radio emission. Instead, we take the conservative 92.2\,mJy $3\sigma$ rms value as the upper limit on any radio flare emission. As before, we plot the parameter space in \autoref{fig:parameters_240414a}. For redshift against \eb, we keep $\beta=1.6$ as a constant (following \citealt{starling2020}). For $\beta$ against \eb, we keep redshift constant at 1.833, as reported in GCNs following this burst \citep{gcn_adami2024,gcn_deugartepostigo2024}. We find an upper limit \eb value of $2.76\times 10^{-4}$ to matches our observations.

        We discuss these limits in the context of our LOFAR campaign, and other literature in Section \ref{sec:lofarsample}.

    \subsection{Limitations}

        Across this study, we have adopted a value of DM $\sim 1200z$ following \citet{lorimer2007}. \citet{macquart2020} and \citet{james2022} discuss the assumptions and limitations of this relation in further detail. Additionally, the populations of fast radio bursts (FRBs), a different type of astrophysical transient, on which these DM measures are based may not reflect the local GRB environment. LGRBs preferentially occur in galaxies with high star formation rates \citep{savaglio2009}, meaning we may expect denser environments around the progenitor. Below a redshift $z\lesssim2$, we may expect the contribution to dispersion due to molecular clouds within the host galaxy to become important. A higher dispersion measure on one hand favours our observations, as the delay $\tau(\nu)$ of radio emissions is proportional to DM, resulting in the later arrival of radio flares and relaxing the time required for a radio observatory to get on target. On the other hand, the expected flux density of radio emission is inversely proportional to DM and so any emission becomes even fainter.

        The results and interpretation here also assumed that other propagation effects are not heavily impacting our observations. Radio emission could be self-absorbed locally, or the plasma co-moving frequency may be sufficiently high that some radio emission is cut off. These would act to reduce the radio flux reaching the observer, meaning we are over-predicting what we are expecting to see -- instead any incoming signals may be lost within the noise of the data, if any escapes the progenitor region at all. Although we cannot quantify these effects, previous works have suggested it can be very small \citep{zhang2014} - see \citet{rowlinson2019a} for an in-depth discussion of low frequency radio emission and the propagation effects that may occur. 
    
        One key limitation in being able to more robustly constrain $\epsilon_{B}$ in the context of our magnetic wind model is the lack of a spectroscopic redshift in many bursts. When a redshift is available, we can predict the timing of radio flares and use flux density measurements across this time period to place constraints on our model, rather than using the rms across the whole observation.

\section{Discussion}
    \label{sec:discussion}

    We have carried out a rapid follow-up search for 144\,MHz coherent radio emission in four long GRBs, finding no evidence of significant emission in any case. In the following sections, we discuss these results in the context of our full LOFAR rapid follow-up campaign, previous literature estimates of the \eb parameter and future prospects of prompt radio follow-up of GRBs.
    
    \subsection{LOFAR rapid follow-up campaign}
        \label{sec:lofarsample}

        \begin{table*}
            \caption{Summary of the sample of LGRBs observed as part of the LOFAR rapid response follow-up campaign. The search for radio emission in GRB\,180706A was investigated through a magnetar central engine model, as opposed to looking for emission analogous to prompt gamma-ray pulses or X-ray flares.}
            \label{tab:allGRBs}
            \begin{tabular}{lcccccccl}
                \hline
                GRB & BAT T90 & $z$ & LOFAR $T_{\textrm{start}}$ (s since & $S_{\nu}$ limit (2h) & \multicolumn{2}{c}{$3\sigma\,S_{\nu}$ time sliced limits} & \eb constraint & References \\
                 & (seconds) &  & BAT trigger) & (mJy) & Bin length (s) & Limit (mJy) &  &  \\ \hline
                \multirow{2}{*}{180706A} & \multirow{2}{*}{$42.7 \pm 8.7$} & \multirow{2}{*}{$\le 2$} & \multirow{2}{*}{260} & \multirow{2}{*}{1.7\,beam$^{-1}$} & 30 & 28\,beam$^{-1}$ & \multirow{2}{*}{$\lesssim [0.24-0.47]$} & \multirow{2}{*}{[1], [2]} \\
                 &  &  &  &  & 120 & 11\,beam$^{-1}$ &  &  \\ \hline
                \multirow{2}{*}{200925B} & \multirow{2}{*}{$18.25 \pm 0.97$} & \multirow{2}{*}{$\sim$1.8} & \multirow{2}{*}{267} & \multirow{2}{*}{3.1} & 60 & 51.3 & \multirow{2}{*}{$< 4.22\times 10^{-4}$} & \multirow{2}{*}{{\it this work}, [3]} \\
                 &  &  &  &  & 180 & 28.8 &  &  \\
                \multirow{2}{*}{210104A} & \multirow{2}{*}{$32.06 \pm 0.49$} & \multirow{2}{*}{0.46} & \multirow{2}{*}{269} & \multirow{2}{*}{3.6} & 60 & 73.5 & \multirow{2}{*}{-} & \multirow{2}{*}{{\it this work}, [4], [5]} \\
                 &  &  &  &  & 180 & 45.3 &  &  \\
                \multirow{2}{*}{210112A} & \multirow{2}{*}{$107.6 \pm 13.0$} & \multirow{2}{*}{$\sim$2} & \multirow{2}{*}{511} & \multirow{2}{*}{3} & 60 & 87.0 & \multirow{2}{*}{$\lesssim 10^{-4}$} & \multirow{2}{*}{[6], [7], [8]} \\
                 &  &  &  &  & 320 & 42.0 &  &  \\
                \multirow{2}{*}{240414A} & \multirow{2}{*}{$88.28 \pm 50.41$} & \multirow{2}{*}{1.833} & \multirow{2}{*}{373} & \multirow{2}{*}{3.1} & 60 & 92.2 & $< 2.76\times 10^{-4}$ & {\it this work}, [9], [10] \\
                 &  &  &  &  & 180 & 65.4 &  &  \\
                \multirow{2}{*}{240418A} & \multirow{2}{*}{$12.00 \pm 3.61$} & \multirow{2}{*}{-} & \multirow{2}{*}{373} & \multirow{2}{*}{4.8} & 60 & 157.0 & \multirow{2}{*}{-} & \multirow{2}{*}{{\it this work}, [11]} \\
                 &  &  &  &  & 180 & 79.5 &  &  \\ \hline
                \end{tabular}
            {\bf References}:
            [1] \citet{rowlinson2019}, [2] \citet{gcn_sakamoto2018}, [3] \citet{gcn_stamatikos2020}, [4] \citet{gcn_palmer2021}, [5] \citet{zhang2022a}, [6] \citet{hennessy2023}, [7] \citet{gcn_stamatikos2021}, [8] \citet{gcn_kann2021}, [9] \citet{gcn_markwardt2024}, [10] \citet{gcn_adami2024}, [11] \citet{gcn_palmer2024}
        \end{table*}

        \begin{figure}
            \includegraphics[width=\linewidth]{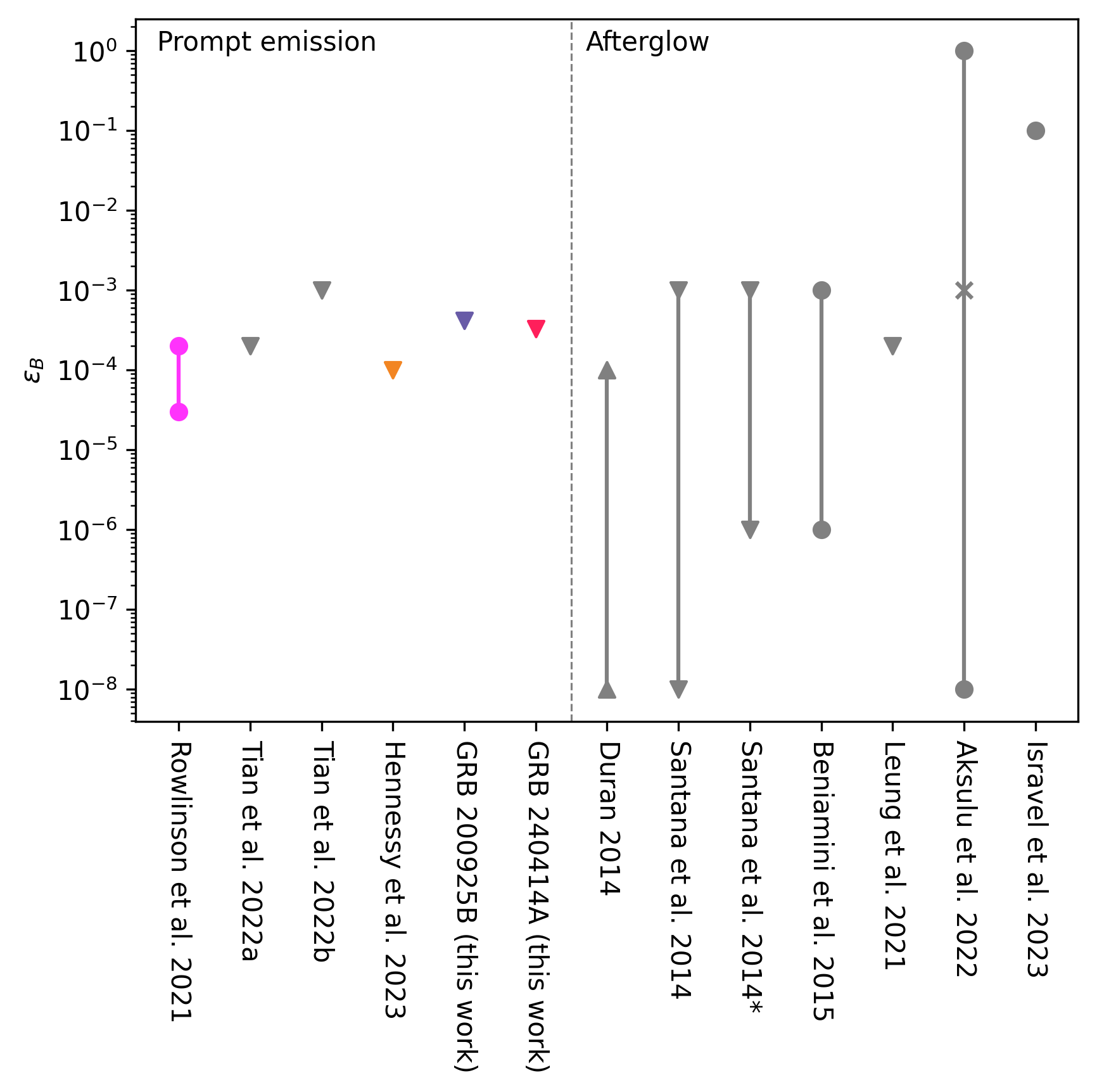}
            \caption{\eb estimates from prompt emission (left of dashed line) and afterglow modelling (right of dashed line) presented from recent literature and this LOFAR sample, sorted in order of publishing year. Up caret are lower limit values, down caret are upper limit values, and circles are measured values.}
            \label{fig:eb_values}
        \end{figure}

        % Sample overview
        Our LOFAR campaign has followed up 6 \swift\ long-duration GRBs and one short GRB over a period of six years, during which the trigger criteria have evolved. This collection of gamma-ray bursts do not therefore form a sample in a statistical sense and are analysed here individually, but act as a pilot for future more homogeneously-selected events. The LGRBs within this sample are summarised in \autoref{tab:allGRBs}. We find that the bursts in this sample are all `standard' in our usual expectations for LGRBs. Temporally, they look to have some or all of the components of a canonical light curve. In the distribution of isotropic energy, luminosity and redshift parameters among the population of \swift\ GRBs, they look average \citep{yi2016}. They also sit comfortably on the Amati relation for LGRBs \citep{amati2006}, a correlation between the peak energy of the prompt burst spectrum and the intrinsic isotropic energy of the burst; and the spectral hardness-T90 distribution among \swift\ bursts \citep{kouveliotou1993,lien2016} showing we are not probing any particularly outlying parameter space.

        A previous study searched for low frequency emission in GRB\,210112A \citep{hennessy2023}, for which a tentative redshift estimate was available, rather than an accurate spectroscopic measurement. In this study, 
        we include two GRBs for which spectroscopically derived redshifts have been reported: GRB\,210104A and GRB\,240414A, which allow us to exactly identify the time window that low frequency pulses are predicted to occur in, and directly probe coherent radio emission related to the energetic processes in the prompt phase. GRB\,240414A has both a robust redshift and clear flaring activity, however the LOFAR observing window catches only the end of the flaring period. GRB\,210104A and GRB\,240418A presented in this paper are blindly searched for any persistent or flaring radio activity, in the former case due to prompt radio activity not being within the observation window, and in the latter due to a lack of redshift and any gamma/X-ray flaring in the \swift\ light curve. GRB\,200925B had no reported redshift and also constitutes a blind search, which recovers a marginal flux density enhancement, discussed in \autoref{sec:discussion_200925b}.

        We did not detect significant signs of radio emission in any case, but this may not be unexpected: the results of this paper are in line with our previous study of GRB\,210112A \citep{hennessy2023}, where we commonly conclude that a lower value of \eb is likely required, where the model consequently produces radio predictions at or lower than typical noise levels within LOFAR images. For the initial model input of $\epsilon_{B} = 10^{-3}$, LOFAR is expected to be more than sensitive enough to have observed most of the predicted radio flares.

        An alternative means of producing coherent radio emission is through a rapidly rotating, magnetised neutron star \citep{rowlinson2019}. Even if only formed for a short duration, this provides a method of producing persistent or pulsed radio emission. As the remnant spins down, emission can occur through dipole magnetic breaking \citep{totani2013}. Some models predict the formation of such objects during a GRB \citep{rowlinson2010,rowlinson2013}. X-ray plateaus seen in some afterglows are a feature associated with these objects \citep{zhang2001,zhang2014}. See \citet[][and references therein]{rowlinson2019a} for a further discussion of the magnetar model. We do not consider this model in this paper, but previous studies have looked for such emission \citep[e.g.][]{curtin2023}, including two bursts followed up by LOFAR. The first LGRB of the LOFAR campaign, GRB\,180706A, was interpreted in this framework of magnetar models. While flares were present in the early \swift\ data, they did not fall in LOFAR's observation window, and so this burst could not be probed for low frequency emission through the magnetic wind model. Instead, the plateau phase is probed, looking for evidence of a magnetar forming and producing persistent or flaring radio activity. No short duration emission is found to deep limits, though this does not rule out the model. More notably, a candidate radio flash was detected and shown to be associated with the short GRB\,201006A \citep{rowlinson2024}. This is shown to be consistent with emission from prolonged central engine activity through the formation of a magnetar during the burst. This was a burst followed-up by LOFAR in rapid response mode, but not included here as it is a short-duration GRB expected to have been formed from a compact binary merger, rather than stellar collapse.

    \subsection{Previous estimates of $\mathbf{\epsilon_{B}}$}
    
        The implications for \eb generated in our study are also in agreement with other recent literature publishing constraints on this parameter, derived through a variety of methods and at a variety of times. This is visualised in \autoref{fig:eb_values} where we show our constraints from our LOFAR sample alongside other literature. These studies look at different phases within the GRB; we may expect varying \eb values as the prompt and afterglow do not necessarily reflect the same mechanisms. The variety of values and their large uncertainties highlight the challenge in understanding the contribution of a magnetic field in a GRB jet.
        
        \citet{santana2014} study the forward shocks of GRBs to obtain \eb values using both an X-ray and an optical sample. By assuming the end of the steep decay phase in \swift\ light curves is the decaying forward shock, they deduce upper limits of $10^{-8} \lesssim \epsilon_{B} \lesssim 10^{-3}$. By assuming the observed optical flux arises due to the forward shock, they derive limits of $10^{-6} \lesssim \epsilon_{B} \lesssim 10^{-3}$, with medians of $\sim 10^{-5}$ for both X-ray and optical derived limits. 
    
        Alternatively, some studies look at afterglows as a means of constraining the structure of the GRB jet and its surrounding environment. In a systematic study of magnetic fields in forward shocks, \citet{barniolduran2014} used the time and flux of the GRB radio afterglow peak to determine \eb, assuming the radio emission is produced by the external forward shock. They find limits ranging $\sim10^{-8}-10^{-4}$, with a median at $\sim10^{-5}$. Using Bayesian inference, \citet{aksulu2022} study a sample of long- and short-GRB afterglow data sets to infer the parameters of the outflow. They find a wide range of values $10^{-8} \lesssim \epsilon_{B} \lesssim 1$, with an average of $\sim 10^{-3}$. \citet{leung2021} studied the late radio afterglow of GRB\,171205A and, in combination with early-time radio observations, showed the progenitor could have originated in a stellar wind medium through the evolution of the radio spectral energy distribution - through this, estimating shock parameters of the burst, including an estimate of $\epsilon_{B} = 2\times 10^{-4}$. \citet{beniamini2015} used GeV and X-ray fluxes to estimate \eb, assuming that the electrons are fast cooling and that the X-ray flux is not suppressed by Synchrotron-Self Compton (SSC) absorption, determining weak magnetic fields and $10^{-6} \lesssim \epsilon_{B} \lesssim 10^{-3}$. In a study of the emission of GRB\,221009A, a hybrid model is used, where synchrotron radiation explains optical to X-ray emissions and a proton-synchrotron process explains the very high energy afterglow - this model requires a significant fraction of energy in the magnetic field \eb $\sim0.1$ \citep{isravel2023}. This is possibly a more unique case where the presence of a highly magnetised environment, for example within a supernova remnant, allows for the presence of TeV emission otherwise suppressed by SSC.
    
        With improvements to radio facilities' response times and sensitivities, some recent studies alongside our LOFAR sample have looked to probe for prompt radio emission. \citet{rowlinson2021} place a constraint of $3\times10^{-5} \lesssim \epsilon_B \lesssim 2\times10^{-4}$ with a LOFAR follow-up of short GRB\,181123B, looking for radio emission predicted to occur in the collapse of a neutron star into a black hole. \citet{tian2022} determine a constraint of $\epsilon_B \lesssim 2\times10^{-4}$ in a very rapid follow-up study of short GRBs, based on a non-detection and limiting sensitivity of MWA with an assumed formation of a magnetar. And in an MWA rapid follow-up of GRB\,210419A, \citet{tian2022a} look for radio emission on minute timescales, finding no detection but placing upper limits of radio emission analogous to X-ray flaring activity resulting in an upper limit of $\epsilon_{B} \lesssim 10^{-3}$. 

    \subsection{Future prospects}

        \begin{figure}
            \includegraphics[width=\linewidth]{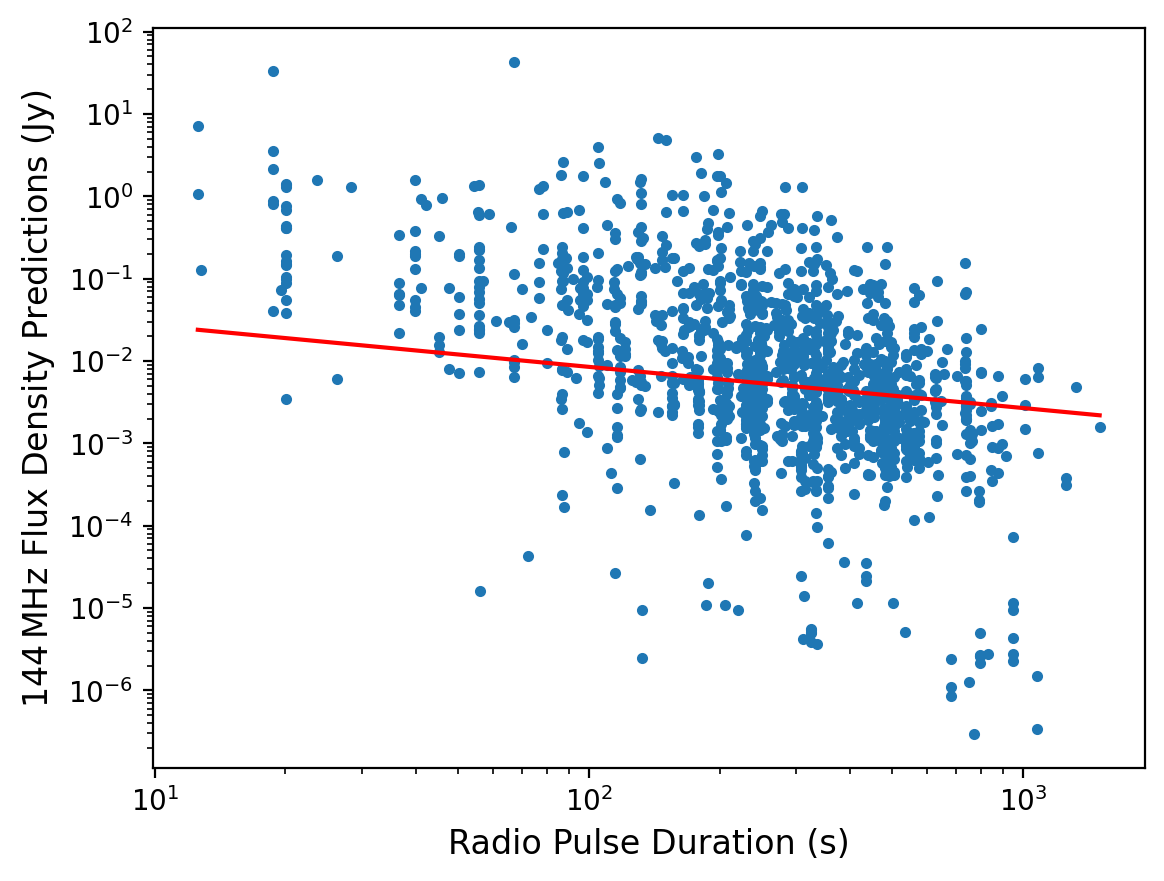}
            \caption{Predictions for the radio flux density for all \swift-BAT pulses with redshift. Above the red line indicates pulses which are detected based on the expected sensitivity and faster response of LOFAR 2.0, and a lower $\epsilon_{B} = 4\times 10^{-4}$ in line with the conclusions of our LOFAR sample study. A 55\% detection fraction is expected in this framework for BAT pulses.}
            \label{fig:pulse_predictions}
        \end{figure}

        The typical LOFAR rapid response of $\sim 5$ minutes in this campaign currently allows us to access gamma-ray pulses or X-ray flares for $z>1$. With the upgrade to LOFAR 2.0\,\footnote{\url{https://www.lofar.eu/wp-content/uploads/2023/04/LOFAR2_0_White_Paper_v2023.1.pdf}}, we should expect a much faster trigger response time and many more triggers. A response time of one minute means that we should expect to observe, accounting for a dispersion delay, all significant radio emission starting from the BAT trigger time for $z>0.25$. Given many gamma-ray pulses and X-ray flares will occur in the first few tens of seconds if not later, this condition is relaxed further. With such a rapid response time, the vast majority of predicted radio flares become observable.
        
        With the prompt gamma-ray emission accessible to LOFAR 2.0, we can determine the expected detection rate of radio pulses directly from gamma-ray pulses, following the method described in \citet{starling2020}. Based on the common conclusions across our LOFAR campaign and magnetic wind model, we calculate these flux density predictions using $\epsilon_{B} = 4\times 10^{-4}$. The sensitivity limit is calculated using equation 5 in \citet{starling2020}, extrapolated from the baseline sensitivity of 1\,mJy for a 2-hour observation with LOFAR 2.0. We obtained the entire catalogue of \swift-BAT GRB prompt emission light curves with redshift from the GRB products available at the UKSSDC\footnote{\url{https://www.swift.ac.uk/xrt_products/}}, and fitted them using a version of the code, described in \citet{hennessy2023}, specifically for BAT prompt emission data. From the output of 1325 individual fitted pulses across this sample, we determine that 55\% of pulses are bright enough to be observable to LOFAR 2.0 for this lower \eb, this is shown by the population of pulses above the red sensitivity line in \autoref{fig:pulse_predictions}. We also estimate the lower limit of \eb that we can probe, assuming all other assumptions hold in this model, finding the detected fraction drops to 1\% for $\epsilon_{B} \sim 5.3 \times 10^{-6}$. Some studies suggest that \eb may be as low as $10^{-8}$ for some bursts. For such a low value, the resultant radio emission in this framework is below the sensitivity of even LOFAR 2.0. If these values are common, this introduces significant uncertainty into the predicted detection rates. Ranges for \eb going down to as low as $10^{-8}$ have been estimated for the afterglow phase, while the prompt phase estimates to date, relevant to rapid radio follow-up, remain above $3 \times 10^{-5}$ \citep{rowlinson2021}.
        
        This promises that current generation radio facilities offer an opportunity to further constrain GRB jets and their environments, utilising the earlier and brighter pulses. A larger set of triggers will allow us to more confidently place constraints on radio emission emitted analogous to prompt emission pulses, especially with a growing number of bursts with redshifts.
    
        The Space Variable Objects Monitor \citep[SVOM,][]{atteia2022} mission has launched in 2024. It can provide GRB triggering in the soft X-ray regime at 4\,keV, lower than other currently operating missions. This means early access to this regime than \swift-XRT can typically provide. Similarly, the Einstein Probe mission \citep[EP,][]{yuan2022} provides a wide-field triggering X-ray telescope, operating at 0.5-4\,keV.

        The Square Kilometre Array Low Frequency Telescope \citep[SKA-Low,][]{dewdney2009} represents the next generation of radio telescope. It will operate in the 50-300\,MHz frequency range and offer deeper sensitivity than current telescopes like LOFAR and MWA. For 60 and 180 second integration times like the images created in this study, we can expect sensitivities\footnote{\url{https://sensitivity-calculator.skao.int/low}} on the order of a few 0.1\,mJy \citep{sokolowski2022}. Compared to the limits of a few tens of mJy in this paper, this is an improvement of two orders of magnitude.
    
        To further probe the mechanics of the GRB inner engine, and the predictions of radio emission analogous to gamma-ray and X-ray flares, we need continued rapid follow-up observations of GRB with radio facilities like LOFAR and MWA, alongside continued \swift\, SVOM, EP, and other high energy GRB triggering facilities. A large sample of bursts is important especially due to the unpredictability of GRB pulses/flares (if any occur at all); uncertainty within the model parameters; and limited knowledge of the environment and its effects on emission - immediately around the burst, and in host galaxies. A large sample provides the ability to fold sample parameter distributions through observed flux limit distributions, to give better statistics. In addition, optical follow-up to provide consistent and accurate spectroscopic redshifts will allow us to probe the model with more confidence.

\section{Conclusion}
    \label{sec:conclusion}

    In this paper, we have presented the \swift\ and LOFAR HBA rapid-response observations of four long gamma-ray bursts. For two of them, we directly test a magnetic wind model, predicting and looking for radio emission emitted analogous to the observed pulses seen at gamma-ray and X-ray wavelengths. A detection in line with predictions would evidence the GRB jet is Poynting flux dominated.

    No significant emission is detected in any of the observations, at either 60\,s or 180\,s timescales. Assuming the \citet{usov2000} magnetic wind model, our constraints suggest the input parameters must be tweaked to account for the measured flux density limits, which are typically low. Primarily, a key unknown is the proportion of internal energy in the magnetic fields behind the shock - our radio predictions are heavily dependent on this value, and based on a non-detection we were able to derive limits on this parameter for GRBs emitting prompt radio emission alongside other more energetic wavelengths.
    
    In GRB\,200925B, we produce $3\sigma$ 144\,MHz flux density upper limits of 51.3 and 28.8\,mJy for 60 and 180 second time slices, respectively. We see a small rise in flux with a low significance of $2\sigma$. If it were real, it falls in line with timing and duration predictions for $z=1.8$ and would imply an \eb upper limit of $4.2\times10^{-4}$ to match our observations. If it were not real, we produce an upper limit of $5.6\times10^{-4}$. In GRB\,240414A, based on 180 second $3\sigma$ 144\,MHz flux density upper limit of 92.2\,mJy we produce an \eb upper limit of $2.8\times10^{-4}$. In 60 second timeslices we find a radio upper limit of 65.4\,mJy. For the remaining two bursts, we could not directly test radio emission from prompt engine activity, but we produce similar 60 and 180 second observed flux density limits of 73.5 and 45.3\,mJy for GRB\,210104A; and 157 and 79.5\,mJy for GRB\,240418A. 

    The weak redshift constraints for some GRBs present a further challenge in constraining the model. Given a redshift for GRB\,200925B we would be able to more concretely accept or rule out the possibility of the small flux density enhancement, and in general a precise redshift means our predictions can be more certain both in terms of timing and energetics of incoming prompt radio emission.

    With the advent of new and enhanced facilities for both triggering and follow-up we will be able to build samples with the capability to more robustly test magnetically-dominated models of GRB jets.

\section*{Acknowledgements}
    The authors would like to thank the anonymous reviewer for their valuable feedback, which has greatly improved this work.
    AH is supported by an STFC Studentship.
    RLCS is supported by a Leverhulme Trust Research Project Grant RPG-2023-240 and acknowledges the ASTRON/JIVE Helena Kluyver visitor programme which supported the early phases of this work.
    AR and IdR acknowledge funding from the Nationale Wetenschapsagenda (NWA) CORTEX grant (NWA.1160.18.316) of the research programme NWA-ORC (Onderzoek op Routes door Consortia) which is (partly) financed by the Dutch Research Council (NWO).
    NRT acknowledges support through STFC Consolidated Grant ST/W000857/1.
    This research used the ALICE High Performance Computing facility at the University of Leicester. This work made use of data supplied by the UK Swift Science Data Centre at the University of Leicester.
    This paper is based (in part) on data obtained with the LOFAR telescope (LOFAR-ERIC) under project codes $LC10\_012$ (PI: Rowlinson), $LC14\_004$ (PI: Starling) $LC15\_013$ (PI: Starling) and $LC20\_021$ (PI: Rowlinson). LOFAR \citep{vanhaarlem2013} is the Low Frequency Array designed and constructed by ASTRON. It has observing, data processing, and data storage facilities in several countries, that are owned by various parties (each with their own funding sources), and that are collectively operated by the LOFAR European Research Infrastructure Consortium (LOFAR-ERIC) under a joint scientific policy. The LOFAR-ERIC resources have benefited from the following recent major funding sources: CNRS-INSU, Observatoire de Paris and Universit\'{e} d'Orl\'{e}ans, France; BMBF, MIWF-NRW, MPG, Germany; Science Foundation Ireland (SFI), Department of Business, Enterprise and Innovation (DBEI), Ireland; NWO, The Netherlands; The Science and Technology Facilities Council, UK; Ministry of Science and Higher Education, Poland.
    For the purpose of open access, the author has applied a Creative Commons Attribution (CC BY) licence to the Author Accepted Manuscript version arising from this submission. 

\section*{Data Availability}

    LOFAR data underlying the findings are openly available in the LOFAR Long-Term Archive (LTA) accessible via \url{https://lta.lofar.eu}.
    Swift data and XRT products are openly available via the UK Swift Science Data Centre (UKSSDC) at \url{www.swift.ac.uk}.
    
%%%%%%%%%%%%%%%%%%%% REFERENCES %%%%%%%%%%%%%%%%%%

\bibliographystyle{mnras}
\bibliography{references}

\begin{thebibliography}{}
\makeatletter
\relax
\def\mn@urlcharsother{\let\do\@makeother \do\$\do\&\do\#\do\^\do\_\do\%\do\~}
\def\mn@doi{\begingroup\mn@urlcharsother \@ifnextchar [ {\mn@doi@} {\mn@doi@[]}}
\def\mn@doi@[#1]#2{\def\@tempa{#1}\ifx\@tempa\@empty \href {http://dx.doi.org/#2} {doi:#2}\else \href {http://dx.doi.org/#2} {#1}\fi \endgroup}
\def\mn@eprint#1#2{\mn@eprint@#1:#2::\@nil}
\def\mn@eprint@arXiv#1{\href {http://arxiv.org/abs/#1} {{\tt arXiv:#1}}}
\def\mn@eprint@dblp#1{\href {http://dblp.uni-trier.de/rec/bibtex/#1.xml} {dblp:#1}}
\def\mn@eprint@#1:#2:#3:#4\@nil{\def\@tempa {#1}\def\@tempb {#2}\def\@tempc {#3}\ifx \@tempc \@empty \let \@tempc \@tempb \let \@tempb \@tempa \fi \ifx \@tempb \@empty \def\@tempb {arXiv}\fi \@ifundefined {mn@eprint@\@tempb}{\@tempb:\@tempc}{\expandafter \expandafter \csname mn@eprint@\@tempb\endcsname \expandafter{\@tempc}}}

\bibitem[\protect\citeauthoryear{Adami, Ilbert, {de la Torre}, {de Ugarte Postigo}, Schneider  \& Le~Floc'h}{Adami et~al.}{2024}]{gcn_adami2024}
Adami C.,  Ilbert O.,  {de la Torre} S.,  {de Ugarte Postigo} A.,  Schneider B.,   Le~Floc'h E.,  2024, GRB Coordinates Network, 36085, 1

\bibitem[\protect\citeauthoryear{Aksulu, Wijers, {van Eerten}  \& {van der Horst}}{Aksulu et~al.}{2022}]{aksulu2022}
Aksulu M.~D.,  Wijers R. A. M.~J.,  {van Eerten} H.~J.,   {van der Horst} A.~J.,  2022, \mn@doi [Monthly Notices of the Royal Astronomical Society] {10.1093/mnras/stac246}, 511, 2848

\bibitem[\protect\citeauthoryear{Amati}{Amati}{2006}]{amati2006}
Amati L.,  2006, \mn@doi [Monthly Notices of the Royal Astronomical Society] {10.1111/j.1365-2966.2006.10840.x}, 372, 233

\bibitem[\protect\citeauthoryear{Anderson et~al.,}{Anderson et~al.}{2018}]{anderson2018}
Anderson G.~E.,  et~al., 2018, \mn@doi [Monthly Notices of the Royal Astronomical Society] {10.1093/mnras/stx2407}, 473, 1512

\bibitem[\protect\citeauthoryear{Anderson et~al.,}{Anderson et~al.}{2021}]{anderson2021}
Anderson G.~E.,  et~al., 2021, \mn@doi [Publications of the Astronomical Society of Australia] {10.1017/pasa.2021.15}, 38, e026

\bibitem[\protect\citeauthoryear{Atteia, Cordier  \& Wei}{Atteia et~al.}{2022}]{atteia2022}
Atteia J.~L.,  Cordier B.,   Wei J.,  2022, \mn@doi [International Journal of Modern Physics D] {10.1142/S0218271822300087}, 31, 2230008

\bibitem[\protect\citeauthoryear{Barniol~Duran}{Barniol~Duran}{2014}]{barniolduran2014}
Barniol~Duran R.,  2014, \mn@doi [Monthly Notices of the Royal Astronomical Society] {10.1093/mnras/stu1070}, 442, 3147

\bibitem[\protect\citeauthoryear{Barthelmy et~al.,}{Barthelmy et~al.}{2005}]{barthelmy2005}
Barthelmy S.~D.,  et~al., 2005, \mn@doi [Space Science Reviews] {10.1007/s11214-005-5096-3}, 120, 143

\bibitem[\protect\citeauthoryear{Beniamini \& Granot}{Beniamini \& Granot}{2016}]{beniamini2016}
Beniamini P.,  Granot J.,  2016, \mn@doi [Monthly Notices of the Royal Astronomical Society] {10.1093/mnras/stw895}, 459, 3635

\bibitem[\protect\citeauthoryear{Beniamini, Nava, Duran  \& Piran}{Beniamini et~al.}{2015}]{beniamini2015}
Beniamini P.,  Nava L.,  Duran R.~B.,   Piran T.,  2015, \mn@doi [Monthly Notices of the Royal Astronomical Society] {10.1093/mnras/stv2033}, 454, 1073

\bibitem[\protect\citeauthoryear{Burrows et~al.,}{Burrows et~al.}{2005}]{burrows2005}
Burrows D.~N.,  et~al., 2005, \mn@doi [Space Science Reviews] {10.1007/s11214-005-5097-2}, 120, 165

\bibitem[\protect\citeauthoryear{Cepa}{Cepa}{1998}]{cepa1998}
Cepa J.,  1998, \mn@doi [Astrophysics and Space Science] {10.1023/A:1002144913887}, 263, 369

\bibitem[\protect\citeauthoryear{Chandra \& Frail}{Chandra \& Frail}{2012}]{chandra2012}
Chandra P.,  Frail D.~A.,  2012, \mn@doi [The Astrophysical Journal] {10.1088/0004-637X/746/2/156}, 746, 156

\bibitem[\protect\citeauthoryear{Curtin et~al.,}{Curtin et~al.}{2023}]{curtin2023}
Curtin A.~P.,  et~al., 2023, \mn@doi [The Astrophysical Journal] {10.3847/1538-4357/ace52f}, 954, 154

\bibitem[\protect\citeauthoryear{Dewdney, Hall, Schilizzi  \& Lazio}{Dewdney et~al.}{2009}]{dewdney2009}
Dewdney P.~E.,  Hall P.~J.,  Schilizzi R.~T.,   Lazio T. J. L.~W.,  2009, \mn@doi [Proceedings of the IEEE] {10.1109/JPROC.2009.2021005}, 97, 1482

\bibitem[\protect\citeauthoryear{Dichiara et~al.,}{Dichiara et~al.}{2024}]{gcn_dichiara2024}
Dichiara S.,  et~al., 2024, GRB Coordinates Network, 36104, 1

\bibitem[\protect\citeauthoryear{Evans et~al.,}{Evans et~al.}{2007}]{evans2007}
Evans P.~A.,  et~al., 2007, \mn@doi [Astronomy \& Astrophysics] {10.1051/0004-6361:20077530}, 469, 379

\bibitem[\protect\citeauthoryear{Evans et~al.,}{Evans et~al.}{2009}]{evans2009}
Evans P.~A.,  et~al., 2009, \mn@doi [Monthly Notices of the Royal Astronomical Society] {10.1111/j.1365-2966.2009.14913.x}, 397, 1177

\bibitem[\protect\citeauthoryear{Evans et~al.,}{Evans et~al.}{2010}]{evans2010}
Evans P.~A.,  et~al., 2010, \mn@doi [Astronomy and Astrophysics] {10.1051/0004-6361/201014819}, 519, A102

\bibitem[\protect\citeauthoryear{Gehrels et~al.,}{Gehrels et~al.}{2004}]{gehrels2004}
Gehrels N.,  et~al., 2004, \mn@doi [The Astrophysical Journal] {10.1086/422091}, 611, 1005

\bibitem[\protect\citeauthoryear{Granot \& van~der Horst}{Granot \& van~der Horst}{2014}]{granot2014}
Granot J.,  van~der Horst A.~J.,  2014, \mn@doi [Publications of the Astronomical Society of Australia] {10.1017/pasa.2013.44}, 31, e008

\bibitem[\protect\citeauthoryear{Hancock et~al.,}{Hancock et~al.}{2019}]{hancock2019}
Hancock P.~J.,  et~al., 2019, \mn@doi [Publications of the Astronomical Society of Australia] {10.1017/pasa.2019.40}, 36, e046

\bibitem[\protect\citeauthoryear{Hennessy et~al.,}{Hennessy et~al.}{2023}]{hennessy2023}
Hennessy A.,  et~al., 2023, \mn@doi [Monthly Notices of the Royal Astronomical Society] {10.1093/mnras/stad2670}, 526, 106

\bibitem[\protect\citeauthoryear{Intema, Jagannathan, Mooley  \& Frail}{Intema et~al.}{2017}]{intema2017}
Intema H.~T.,  Jagannathan P.,  Mooley K.~P.,   Frail D.~A.,  2017, \mn@doi [Astronomy and Astrophysics] {10.1051/0004-6361/201628536}, 598, A78

\bibitem[\protect\citeauthoryear{Isravel, B{\'e}gu{\'e}  \& Pe'er}{Isravel et~al.}{2023}]{isravel2023}
Isravel H.,  B{\'e}gu{\'e} D.,   Pe'er A.,  2023, \mn@doi [The Astrophysical Journal] {10.3847/1538-4357/acefcd}, 956, 12

\bibitem[\protect\citeauthoryear{James, Prochaska, Macquart, {North-Hickey}, Bannister  \& Dunning}{James et~al.}{2022}]{james2022}
James C.~W.,  Prochaska J.~X.,  Macquart J.-P.,  {North-Hickey} F.~O.,  Bannister K.~W.,   Dunning A.,  2022, \mn@doi [Monthly Notices of the Royal Astronomical Society] {10.1093/mnras/stab3051}, 509, 4775

\bibitem[\protect\citeauthoryear{Kann et~al.,}{Kann et~al.}{2011}]{kann2011}
Kann D.~A.,  et~al., 2011, \mn@doi [The Astrophysical Journal] {10.1088/0004-637X/734/2/96}, 734, 96

\bibitem[\protect\citeauthoryear{Kann, {de Ugarte Postigo}, Thoene, Blazek, Agui~Fernandez  \& Sota}{Kann et~al.}{2021}]{gcn_kann2021}
Kann D.~A.,  {de Ugarte Postigo} A.,  Thoene C.~C.,  Blazek M.,  Agui~Fernandez J.~F.,   Sota A.,  2021, GRB Coordinates Network, 29296, 1

\bibitem[\protect\citeauthoryear{Karambelkar et~al.,}{Karambelkar et~al.}{2024}]{gcn_karambelkar2024}
Karambelkar V.,  et~al., 2024, GRB Coordinates Network, 36199, 1

\bibitem[\protect\citeauthoryear{Katz}{Katz}{1997}]{katz1997}
Katz J.~I.,  1997, \mn@doi [The Astrophysical Journal] {10.1086/304896}, 490, 633

\bibitem[\protect\citeauthoryear{Kennea et~al.,}{Kennea et~al.}{2021}]{gcn_kennea2021}
Kennea J.~A.,  et~al., 2021, GRB Coordinates Network, 29260, 1

\bibitem[\protect\citeauthoryear{Klebesadel, Strong  \& Olson}{Klebesadel et~al.}{1973}]{klebesadel1973}
Klebesadel R.~W.,  Strong I.~B.,   Olson R.~A.,  1973, \mn@doi [The Astrophysical Journal] {10.1086/181225}, 182, L85

\bibitem[\protect\citeauthoryear{Kosogorov et~al.,}{Kosogorov et~al.}{2025}]{kosogorov2025}
Kosogorov N.,  et~al., 2025, \mn@doi [The Astrophysical Journal] {10.3847/1538-4357/add014}, 985, 265

\bibitem[\protect\citeauthoryear{Kouveliotou, Meegan, Fishman, Bhat, Briggs, Koshut, Paciesas  \& Pendleton}{Kouveliotou et~al.}{1993}]{kouveliotou1993}
Kouveliotou C.,  Meegan C.~A.,  Fishman G.~J.,  Bhat N.~P.,  Briggs M.~S.,  Koshut T.~M.,  Paciesas W.~S.,   Pendleton G.~N.,  1993, \mn@doi [The Astrophysical Journal Letters] {10.1086/186969}, 413, L101

\bibitem[\protect\citeauthoryear{Kumar \& Zhang}{Kumar \& Zhang}{2015}]{kumar2015}
Kumar P.,  Zhang B.,  2015, \mn@doi [Physics Reports] {10.1016/j.physrep.2014.09.008}, 561, 1

\bibitem[\protect\citeauthoryear{Lan, Wei, Zeng, Li  \& Wu}{Lan et~al.}{2021}]{lan2021}
Lan G.-X.,  Wei J.-J.,  Zeng H.-D.,  Li Y.,   Wu X.-F.,  2021, \mn@doi [Monthly Notices of the Royal Astronomical Society] {10.1093/mnras/stab2508}, 508, 52

\bibitem[\protect\citeauthoryear{Lazzati, Morsony, Margutti  \& Begelman}{Lazzati et~al.}{2013}]{lazzati2013}
Lazzati D.,  Morsony B.~J.,  Margutti R.,   Begelman M.~C.,  2013, \mn@doi [The Astrophysical Journal] {10.1088/0004-637X/765/2/103}, 765, 103

\bibitem[\protect\citeauthoryear{Leung et~al.,}{Leung et~al.}{2021}]{leung2021}
Leung J.~K.,  et~al., 2021, \mn@doi [Monthly Notices of the Royal Astronomical Society] {10.1093/mnras/stab326}, 503, 1847

\bibitem[\protect\citeauthoryear{Lien et~al.,}{Lien et~al.}{2016}]{lien2016}
Lien A.,  et~al., 2016, \mn@doi [The Astrophysical Journal] {10.3847/0004-637X/829/1/7}, 829, 7

\bibitem[\protect\citeauthoryear{Lorimer, Bailes, McLaughlin, Narkevic  \& Crawford}{Lorimer et~al.}{2007}]{lorimer2007}
Lorimer D.~R.,  Bailes M.,  McLaughlin M.~A.,  Narkevic D.~J.,   Crawford F.,  2007, \mn@doi [Science] {10.1126/science.1147532}, 318, 777

\bibitem[\protect\citeauthoryear{Ma, Xie, Wang, Dong  \& Zhi}{Ma et~al.}{2024}]{ma2024}
Ma G.-L.,  Xie W.,  Wang W.-K.,  Dong A.-J.,   Zhi Q.-J.,  2024, \mn@doi [Astronomische Nachrichten] {10.1002/asna.20230179}, 345, e20230179

\bibitem[\protect\citeauthoryear{Macquart et~al.,}{Macquart et~al.}{2020}]{macquart2020}
Macquart J.~P.,  et~al., 2020, \mn@doi [Nature] {10.1038/s41586-020-2300-2}, 581, 391

\bibitem[\protect\citeauthoryear{Markwardt et~al.,}{Markwardt et~al.}{2024}]{gcn_markwardt2024}
Markwardt C.~B.,  et~al., 2024, GRB Coordinates Network, 36140, 1

\bibitem[\protect\citeauthoryear{Norris, Nemiroff, Bonnell, Scargle, Kouveliotou, Paciesas, Meegan  \& Fishman}{Norris et~al.}{1996}]{norris1996}
Norris J.~P.,  Nemiroff R.~J.,  Bonnell J.~T.,  Scargle J.~D.,  Kouveliotou C.,  Paciesas W.~S.,  Meegan C.~A.,   Fishman G.~J.,  1996, \mn@doi [The Astrophysical Journal] {10.1086/176902}, 459, 393

\bibitem[\protect\citeauthoryear{Offringa, {de Bruyn}, Biehl, Zaroubi, Bernardi  \& Pandey}{Offringa et~al.}{2010}]{offringa2010}
Offringa A.~R.,  {de Bruyn} A.~G.,  Biehl M.,  Zaroubi S.,  Bernardi G.,   Pandey V.~N.,  2010, \mn@doi [Monthly Notices of the Royal Astronomical Society] {10.1111/j.1365-2966.2010.16471.x}, 405, 155

\bibitem[\protect\citeauthoryear{Offringa, {van de Gronde}  \& Roerdink}{Offringa et~al.}{2012}]{offringa2012}
Offringa A.~R.,  {van de Gronde} J.~J.,   Roerdink J. B. T.~M.,  2012, \mn@doi [Astronomy and Astrophysics] {10.1051/0004-6361/201118497}, 539, A95

\bibitem[\protect\citeauthoryear{Offringa et~al.,}{Offringa et~al.}{2014}]{offringa2014}
Offringa A.~R.,  et~al., 2014, \mn@doi [Monthly Notices of the Royal Astronomical Society] {10.1093/mnras/stu1368}, 444, 606

\bibitem[\protect\citeauthoryear{Palmer et~al.,}{Palmer et~al.}{2021}]{gcn_palmer2021}
Palmer D.~M.,  et~al., 2021, GRB Coordinates Network, 29251, 1

\bibitem[\protect\citeauthoryear{Palmer et~al.,}{Palmer et~al.}{2024}]{gcn_palmer2024}
Palmer D.~M.,  et~al., 2024, GRB Coordinates Network, 36254, 1

\bibitem[\protect\citeauthoryear{Pe'er, M{\'e}sz{\'a}ros  \& Rees}{Pe'er et~al.}{2007}]{peer2007}
Pe'er A.,  M{\'e}sz{\'a}ros P.,   Rees M.~J.,  2007, \mn@doi [Philosophical Transactions of the Royal Society of London Series A] {10.1098/rsta.2006.1986}, 365, 1171

\bibitem[\protect\citeauthoryear{Rees \& Meszaros}{Rees \& Meszaros}{1994}]{rees1994}
Rees M.~J.,  Meszaros P.,  1994, \mn@doi [The Astrophysical Journal] {10.1086/187446}, 430, L93

\bibitem[\protect\citeauthoryear{Rhodes et~al.,}{Rhodes et~al.}{2024}]{rhodes2024}
Rhodes L.,  et~al., 2024, \mn@doi [Monthly Notices of the Royal Astronomical Society] {10.1093/mnras/stae2050}, 533, 4435

\bibitem[\protect\citeauthoryear{Ror, Gupta, Aryan, Pandey, Oates, {Castro-Tirado}  \& Kumar}{Ror et~al.}{2024}]{ror2024}
Ror A.~K.,  Gupta R.,  Aryan A.,  Pandey S.~B.,  Oates S.~R.,  {Castro-Tirado} A.~J.,   Kumar S.,  2024, \mn@doi [The Astrophysical Journal] {10.3847/1538-4357/ad5554}, 971, 163

\bibitem[\protect\citeauthoryear{Rowlinson \& Anderson}{Rowlinson \& Anderson}{2019}]{rowlinson2019a}
Rowlinson A.,  Anderson G.~E.,  2019, \mn@doi [Monthly Notices of the Royal Astronomical Society] {10.1093/mnras/stz2295}, 489, 3316

\bibitem[\protect\citeauthoryear{Rowlinson et~al.,}{Rowlinson et~al.}{2010}]{rowlinson2010}
Rowlinson A.,  et~al., 2010, \mn@doi [Monthly Notices of the Royal Astronomical Society] {10.1111/j.1365-2966.2010.17354.x}, 409, 531

\bibitem[\protect\citeauthoryear{Rowlinson, O'Brien, Metzger, Tanvir  \& Levan}{Rowlinson et~al.}{2013}]{rowlinson2013}
Rowlinson A.,  O'Brien P.~T.,  Metzger B.~D.,  Tanvir N.~R.,   Levan A.~J.,  2013, \mn@doi [Monthly Notices of the Royal Astronomical Society] {10.1093/mnras/sts683}, 430, 1061

\bibitem[\protect\citeauthoryear{Rowlinson et~al.,}{Rowlinson et~al.}{2019}]{rowlinson2019}
Rowlinson A.,  et~al., 2019, \mn@doi [Monthly Notices of the Royal Astronomical Society] {10.1093/mnras/stz2866}, 490, 3483

\bibitem[\protect\citeauthoryear{Rowlinson et~al.,}{Rowlinson et~al.}{2021}]{rowlinson2021}
Rowlinson A.,  et~al., 2021, \mn@doi [Monthly Notices of the Royal Astronomical Society] {10.1093/mnras/stab2060}, 506, 5268

\bibitem[\protect\citeauthoryear{Rowlinson et~al.,}{Rowlinson et~al.}{2024}]{rowlinson2024}
Rowlinson A.,  et~al., 2024, \mn@doi [Monthly Notices of the Royal Astronomical Society] {10.1093/mnras/stae2234}, p. stae2234

\bibitem[\protect\citeauthoryear{Ryde et~al.,}{Ryde et~al.}{2010}]{ryde2010}
Ryde F.,  et~al., 2010, \mn@doi [The Astrophysical Journal] {10.1088/2041-8205/709/2/L172}, 709, L172

\bibitem[\protect\citeauthoryear{Sakamoto et~al.,}{Sakamoto et~al.}{2018}]{gcn_sakamoto2018}
Sakamoto T.,  et~al., 2018, GRB Coordinates Network, 22921, 1

\bibitem[\protect\citeauthoryear{Santana, Barniol~Duran  \& Kumar}{Santana et~al.}{2014}]{santana2014}
Santana R.,  Barniol~Duran R.,   Kumar P.,  2014, \mn@doi [The Astrophysical Journal] {10.1088/0004-637X/785/1/29}, 785, 29

\bibitem[\protect\citeauthoryear{Sari \& Piran}{Sari \& Piran}{1997}]{sari1997a}
Sari R.,  Piran T.,  1997, \mn@doi [Monthly Notices of the Royal Astronomical Society] {10.1093/mnras/287.1.110}, 287, 110

\bibitem[\protect\citeauthoryear{Savaglio, Glazebrook  \& Le~Borgne}{Savaglio et~al.}{2009}]{savaglio2009}
Savaglio S.,  Glazebrook K.,   Le~Borgne D.,  2009, \mn@doi [The Astrophysical Journal] {10.1088/0004-637X/691/1/182}, 691, 182

\bibitem[\protect\citeauthoryear{Sbarufatti et~al.,}{Sbarufatti et~al.}{2020}]{gcn_sbarufatti2020}
Sbarufatti B.,  et~al., 2020, GRB Coordinates Network, 28508, 1

\bibitem[\protect\citeauthoryear{Schmitt et~al.,}{Schmitt et~al.}{2024}]{schmitt2024}
Schmitt J.,  et~al., 2024, \mn@doi [Astronomy and Astrophysics] {10.1051/0004-6361/202449254}, 687, A198

\bibitem[\protect\citeauthoryear{Smolsky \& Usov}{Smolsky \& Usov}{2000}]{smolsky2000}
Smolsky M.~V.,  Usov V.~V.,  2000, \mn@doi [The Astrophysical Journal] {10.1086/308488}, 531, 764

\bibitem[\protect\citeauthoryear{Sokolowski et~al.,}{Sokolowski et~al.}{2022}]{sokolowski2022}
Sokolowski M.,  et~al., 2022, \mn@doi [Publications of the Astronomical Society of Australia] {10.1017/pasa.2021.63}, 39, e015

\bibitem[\protect\citeauthoryear{Staley}{Staley}{2014}]{staley2014}
Staley T.~D.,  2014, Astrophysics Source Code Library, p. ascl:1411.003

\bibitem[\protect\citeauthoryear{Stamatikos et~al.,}{Stamatikos et~al.}{2020}]{gcn_stamatikos2020}
Stamatikos M.,  et~al., 2020, GRB Coordinates Network, 28506, 1

\bibitem[\protect\citeauthoryear{Stamatikos et~al.,}{Stamatikos et~al.}{2021}]{gcn_stamatikos2021}
Stamatikos M.,  et~al., 2021, GRB Coordinates Network, 29297, 1

\bibitem[\protect\citeauthoryear{Starling, Rowlinson, {van~der~Horst}  \& Wijers}{Starling et~al.}{2020}]{starling2020}
Starling R. L.~C.,  Rowlinson A.,  {van~der~Horst} A.~J.,   Wijers R. A. M.~J.,  2020, \mn@doi [Monthly Notices of the Royal Astronomical Society] {10.1093/mnras/staa1168}, 494, 5787

\bibitem[\protect\citeauthoryear{Swinbank et~al.,}{Swinbank et~al.}{2015}]{LINC}
Swinbank J.~D.,  et~al., 2015, \mn@doi [Astronomy and Computing] {10.1016/j.ascom.2015.03.002}, 11, 25

\bibitem[\protect\citeauthoryear{Taylor \& Cordes}{Taylor \& Cordes}{1993}]{taylor1993}
Taylor J.~H.,  Cordes J.~M.,  1993, \mn@doi [The Astrophysical Journal] {10.1086/172870}, 411, 674

\bibitem[\protect\citeauthoryear{Tian et~al.,}{Tian et~al.}{2022a}]{tian2022}
Tian J.,  et~al., 2022a, \mn@doi [Publications of the Astronomical Society of Australia] {10.1017/pasa.2021.58}, 39, e003

\bibitem[\protect\citeauthoryear{Tian et~al.,}{Tian et~al.}{2022b}]{tian2022a}
Tian J.,  et~al., 2022b, \mn@doi [Monthly Notices of the Royal Astronomical Society] {10.1093/mnras/stac1483}, 514, 2756

\bibitem[\protect\citeauthoryear{Tingay et~al.,}{Tingay et~al.}{2013}]{tingay2013}
Tingay S.~J.,  et~al., 2013, \mn@doi [Publications of the Astronomical Society of Australia] {10.1017/pasa.2012.007}, 30, e007

\bibitem[\protect\citeauthoryear{Tohuvavohu et~al.,}{Tohuvavohu et~al.}{2024}]{gcn_tohuvavohu2024}
Tohuvavohu A.,  et~al., 2024, GRB Coordinates Network, 36181, 1

\bibitem[\protect\citeauthoryear{Totani}{Totani}{2013}]{totani2013}
Totani T.,  2013, \mn@doi [Publications of the Astronomical Society of Japan] {10.1093/pasj/65.5.L12}, 65, L12

\bibitem[\protect\citeauthoryear{Usov}{Usov}{1992}]{usov1992}
Usov V.~V.,  1992, \mn@doi [Nature] {10.1038/357472a0}, 357, 472

\bibitem[\protect\citeauthoryear{Usov \& Katz}{Usov \& Katz}{2000}]{usov2000}
Usov V.~V.,  Katz J.~I.,  2000, \mn@doi [Astronomy and Astrophysics] {10.48550/arXiv.astro-ph/0002278}, 364, 655

\bibitem[\protect\citeauthoryear{Waxman, Kulkarni  \& Frail}{Waxman et~al.}{1998}]{waxman1998}
Waxman E.,  Kulkarni S.~R.,   Frail D.~A.,  1998, \mn@doi [The Astrophysical Journal] {10.1086/305467}, 497, 288

\bibitem[\protect\citeauthoryear{Williams et~al.,}{Williams et~al.}{2016}]{williams2016}
Williams W.~L.,  et~al., 2016, \mn@doi [Monthly Notices of the Royal Astronomical Society] {10.1093/mnras/stw1056}, 460, 2385

\bibitem[\protect\citeauthoryear{Yi, Xi, Yu, Wang, Mu, L{\"u}  \& Liang}{Yi et~al.}{2016}]{yi2016}
Yi S.-X.,  Xi S.-Q.,  Yu H.,  Wang F.~Y.,  Mu H.-J.,  L{\"u} L.-Z.,   Liang E.-W.,  2016, \mn@doi [The Astrophysical Journal Supplement Series] {10.3847/0067-0049/224/2/20}, 224, 20

\bibitem[\protect\citeauthoryear{Yuan, Zhang, Chen  \& Ling}{Yuan et~al.}{2022}]{yuan2022}
Yuan W.,  Zhang C.,  Chen Y.,   Ling Z.,  2022, \mn@doi [Handbook of X-ray and Gamma-ray Astrophysics] {10.1007/978-981-16-4544-0_151-1}, p.~86

\bibitem[\protect\citeauthoryear{Zhang}{Zhang}{2014}]{zhang2014}
Zhang B.,  2014, \mn@doi [The Astrophysical Journal] {10.1088/2041-8205/780/2/L21}, 780, L21

\bibitem[\protect\citeauthoryear{Zhang \& M{\'e}sz{\'a}ros}{Zhang \& M{\'e}sz{\'a}ros}{2001}]{zhang2001}
Zhang B.,  M{\'e}sz{\'a}ros P.,  2001, \mn@doi [The Astrophysical Journal] {10.1086/320255}, 552, L35

\bibitem[\protect\citeauthoryear{Zhang et~al.,}{Zhang et~al.}{2022}]{zhang2022a}
Zhang L.-L.,  et~al., 2022, \mn@doi [The Astrophysical Journal] {10.3847/1538-4357/aca08f}, 941, 63

\bibitem[\protect\citeauthoryear{{de Gasperin} et~al.,}{{de Gasperin} et~al.}{2019}]{degasperin2019}
{de Gasperin} F.,  et~al., 2019, \mn@doi [Astronomy and Astrophysics] {10.1051/0004-6361/201833867}, 622, A5

\bibitem[\protect\citeauthoryear{{de Ugarte Postigo}, Kann, Thoene, Blazek, Agui~Fernandez  \& Aceituno}{{de Ugarte Postigo} et~al.}{2020}]{gcn_deugartepostigo2020}
{de Ugarte Postigo} A.,  Kann D.~A.,  Thoene C.~C.,  Blazek M.,  Agui~Fernandez J.~F.,   Aceituno F.,  2020, GRB Coordinates Network, 28501, 1

\bibitem[\protect\citeauthoryear{{de Ugarte Postigo} et~al.,}{{de Ugarte Postigo} et~al.}{2024}]{gcn_deugartepostigo2024}
{de Ugarte Postigo} A.,  et~al., 2024, GRB Coordinates Network, 36087, 1

\bibitem[\protect\citeauthoryear{{van Haarlem} et~al.,}{{van Haarlem} et~al.}{2013}]{vanhaarlem2013}
{van Haarlem} M.~P.,  et~al., 2013, \mn@doi [Astronomy \&; Astrophysics, Volume 556, id.A2, 53 pp.] {10.1051/0004-6361/201220873}, 556, A2

\bibitem[\protect\citeauthoryear{{van Weeren} et~al.,}{{van Weeren} et~al.}{2016}]{vanweeren2016}
{van Weeren} R.~J.,  et~al., 2016, \mn@doi [The Astrophysical Journal Supplement Series] {10.3847/0067-0049/223/1/2}, 223, 2

\bibitem[\protect\citeauthoryear{van~der Horst et~al.,}{van~der Horst et~al.}{2008}]{horst2008}
van~der Horst A.~J.,  et~al., 2008, \mn@doi [Astronomy \& Astrophysics] {10.1051/0004-6361:20078051}, 480, 35

\makeatother
\end{thebibliography}

%%%%%%%%%%%%%%%%% APPENDICES %%%%%%%%%%%%%%%%%%%%%

% \appendix

%%%%%%%%%%%%%%%%%%%%%%%%%%%%%%%%%%%%%%%%%%%%%%%%%%
% Don't change these lines
\bsp	% typesetting comment
\label{lastpage}
\end{document}